\documentclass[twocolumn,tighten,apj,iop]{emulateapj}
\usepackage[normalem]{ulem}

\newcommand{\pd}[2]{\frac{\partial #1}{\partial #2}}
\newcommand{\td}[2]{\frac{d #1}{d #2}}

\newcommand{\thd}[3]{\biggl(\pd{#1}{#2}\biggr)_{#3}}

\newcommand{\sci}[2]{#1\times 10^{#2}}

\newcommand{\be}{\begin{equation}}
\newcommand{\ee}{\end{equation}}
\newcommand{\bea}{\begin{eqnarray}}
\newcommand{\eea}{\end{eqnarray}}
\newcommand{\ba}{\begin{array}}
\newcommand{\ea}{\end{array}}

\long\def\symbolfootnote[#1]#2{\begingroup%
\def\thefootnote{\fnsymbol{footnote}}\footnote[#1]{#2}\endgroup}

\shortauthors{L. F. Roberts}
\shorttitle{Proto-Neutron Star Evolution}

\begin{document} 

\title{A New Code for Proto-Neutron Star Evolution}
\author{L. F. Roberts$^\dagger$}
\affil{Department of Astronomy and Astrophysics, University of 
California, Santa Cruz, California 95064, USA}
\email{$^\dagger$lroberts@ucolick.org}

\begin{abstract}
A new code for following the evolution and emissions of
proto-neutron stars during the first minute of their lives is
developed and tested.  The code is one dimensional,
fully implicit, and general relativistic. Multi-group, multi-flavor
neutrino transport is incorporated that makes use of variable
Eddington factors obtained from a formal solution of the static
general relativistic Boltzmann equation with linearized scattering
terms.  The timescales of neutrino emission and spectral evolution
obtained using the new code are broadly consistent with previous
results. Unlike other recent calculations, however, the new code predicts
that the neutrino-driven wind will be characterized, at least for part of its 
existence, by a neutron excess.  This change, potentially consequential 
for nucleosynthesis in the wind, is due to an improved
treatment of the charged-current interactions of electron flavored
neutrinos and anti-neutrinos with nucleons.  A comparison is also made 
between the results obtained using either variable Eddington factors or 
simple equilibrium flux-limited diffusion. The latter approximation, which 
has been frequently used in previous studies of proto-neutron star cooling,
accurately describes the total neutrino luminosities (to within 10\%)
for most of the evolution, until the proto-neutron star becomes
optically thin.
\end{abstract}


\maketitle 


\section{Introduction}

A proto-neutron star (PNS) is born after the core of a massive star
collapses to supra-nuclear densities, experiences core bounce
due to the repulsive portion of the nuclear interaction which
launches a shock wave that may eventually serve to disrupt the
entire star in a supernova, and leaves behind a compact remnant.
The overlying star is ejected and some portion of the mass may or
may not fall back \citep[c.f.][]{Janka07}.  In reality the mass of the 
PNS may increase with time due to this accretion, but a frequent assumption 
that is reasonable for low mass progenitors, and
one adopted here, is that the PNS evolves in isolation after the shock
has exited.  Because of the large release of
gravitational binding energy ($2-5 \times10^{53}$ ergs), the PNS is
initially hot and extended compared to a cold neutron star but large
portions of the mass are still at supra-nuclear densities.
Due to the high density and reasonably large temperature of this
nuclear material, it is opaque to neutrinos of all flavors.  In the
outer regions of the PNS where the density is lower, the material is
semi-transparent to neutrinos.  This hot extended object undergoes
Kelvin-Helmholtz cooling by emitting neutrinos of all flavors over a
period of up to a minute \citep{Burrows86}, at which time it
transitions to a phase of optically thin neutrino cooling.

This qualitative description was confirmed when about twenty neutrinos
were observed from SN 1987A \citep{Bionta87,Hirata87}. There were not
enough events, however, to determine much detail of the cooling
process \citep{Lattimer89,Loredo02}, though limits were placed on the
properties of weakly interacting particles \citep{Keil97}.  If a
similar core collapse supernova were to occur today, modern neutrino
detectors would see thousands of events.  Detailed modeling of the
neutrino emission is needed if we are to learn about the central
engine of core collapse supernovae from a nearby event.

There are numerous other reasons why understanding the properties of
late-time supernova neutrinos is important, despite the rarity with
which they are {\it directly} detected.  An important one is the
impact of neutrinos on supernova nucleosynthesis.  Charged current
neutrino interactions in the wind blown from the surface of PNSs
determine the electron fraction of the ejected material and thereby
constrain its nucleosynthesis.  Current uncertainties in the relative
energies of the electron neutrinos and anti-neutrinos are large
enough to allow for both neutron-rich and proton-rich ejecta, which
may be favorable for $r$-process \citep{Woosley94} and $\nu p$-process
nucleosynthesis \citep{Froehlich06,Pruet06}, respectively.  Recent
work points to the wind ejecta being proton rich at all times
\citep{Huedepohl10,Fischer10}, but the reasons for this
change are only beginning to be understood \citep{Fischer11}.
Additionally, the average energies of $\mu$ and $\tau$ neutrinos also
significantly affect the neutrino spallation rates that determine
nucleosynthetic yields of the $\nu$-process \citep{Woosley90}, which
may be responsible for a number of rare isotopes.

It is also possible that current neutrino detectors with upgrades or
next generation neutrino detectors will be able to observe the diffuse
background of neutrinos produced by supernovae over the lifetime of
the universe \citep{Horiuchi09}.  Predictions for the diffuse MeV
scale neutrino background density depend significantly on the
integrated spectrum of neutrinos emitted in core-collapse supernovae
\citep{Woosley86,Ando04}.  The integrated neutrino emission is
dominated by PNS evolution, so that accurate modeling of PNSs
can contribute to understanding the diffuse supernova neutrino
background.
 
Finally, the neutrino emission from the ``photosphere'' of PNSs gives
the initial conditions for the study of both matter-induced and
neutrino-induced neutrino oscillations \citep{Duan06}.  The
differences between the spectra of various neutrino flavors, especially
$\bar \nu_e$ and $\bar \nu_{\mu,\tau}$, can significantly affect the
impact of flavor evolution in the nearly free streaming regime
\citep{Keil03}. The rate of PNS cooling also has the potential to put
limits on exotic physics, such as axions \citep{Keil95b}, the presence
of quark matter or a Kaon condensate in the core
\citep{Pons01a,Pons01b}, as well as possible extensions of the
standard model using data already in hand from SN 1987A.

Theoretical predictions of post-bounce neutrinos have existed for more than 25 
years \citep{Burrows86,Mayle87,Keil95a,Sumiyoshi95,Pons99,Fischer10,Huedepohl10,Roberts12}.  
Since the evolution of PNSs is described by the Kelvin-Helmholtz cooling of the 
collapsed, shock heated remnant of a core-collapse supernova, it is fundamentally 
a radiation hydrodynamics problem (although the regions important for neutrino 
emission are not very {\em dynamic} after bounce).  Over time, the treatment
of radiative transfer and neutrino microphysics in simulations has become 
increasingly sophisticated, moving from the equilibrium flux limited diffusion
(EFLD) and greatly simplified neutrino physics \citep{Burrows86} to full 
solutions of the Boltzmann equation \citep{Fischer10} with more realistic 
microphysics \citep{Huedepohl10}.

Here, a new fully implicit code is developed for calculating the
detailed evolution of PNSs in spherically symmetric general relativity
within a variable Eddington factor formalism.  The structure of the
paper is as follows: In section \ref{sec:RadiativeTransfer}, the
equations of neutrino transport within the projected symmetric trace-free 
moment formalism of
\cite{Thorne81} are described, and generic neutrino source terms for
this formalism are derived.  In section \ref{sec:FormalSolution}, a
method for obtaining closure relations for the moment equations via a
formal solution of the Boltzmann equation are described.  A fully
implicit numerical implementation of neutrino transport coupled to
hydrodynamics/hydrostatics is described in section \ref{sec:Numerics}
(with code tests described in appendix \ref{sec:CodeTests}).  A
fiducial model of PNS cooling is detailed in section
\ref{sec:PNSEvolution}.  These results are compared with the results
of an EFLD calculation of PNS cooling in section \ref{sec:EFLDComp}.
The implications of these new calculations of PNS cooling on the
composition of the neutrino driven wind are discussed in section
\ref{sec:Spectra Compare}.  In section \ref{sec:Integrated Spectra},
the properties of the integrated neutrino emission are discussed.  The
convention $\hbar = c = G = 1$ is adopted in sections
\ref{sec:RadiativeTransfer} through \ref{sec:Numerics} to avoid a
plethora of factors.  In section \ref{sec:Spectra Compare}, units with
$\hbar = c = 1$ are used.

\section{The Moment Approach to General Relativistic Radiative Transfer}
\label{sec:RadiativeTransfer} 
The equations of radiative transfer in curved space-times were first 
derived by \cite{Lindquist66}, which described the evolution of the 
invariant distribution function along geodesics in phase space.  The 
general form of the general relativistic Boltzmann (or Lindquist) 
equation in the absence of external forces is
\be
\label{eq:Boltzmann}
\frac{d f\left(x^\mu,p^\nu(x^\mu)\right)}{d\tau} = 
p^\beta \left(\pd{f}{x^\beta} - \Gamma^{\alpha}_{\beta \gamma} p^\gamma
\pd{f}{p^\alpha} \right) = \left( \td{f}{\tau} \right)_{\rm{coll}},
\ee  
where $f$ is the invariant distribution function, $p^\beta$ is the 
neutrino four-momentum (which is constrained to be on mass shell), 
and $\Gamma^{\alpha}_{\beta \gamma}$ are the Christoffel symbols.  
The collision term on the right hand side describes the destruction and 
production of neutrinos on a particular phase-space trajectory by 
capture processes, pair annihilation, scattering, and their inverses. 
In addition to describing the propagation of neutrinos along trajectories
in physical space, this also encodes the evolution of the energy of 
neutrinos along geodesics of the spacetime.  In three spatial dimensions, 
this is a seven dimensional equation that needs to be solved for each 
neutrino species.  

A number of numerical strategies can be employed to solve the
transport problem \citep{Mihalas84}.  Foremost among these are discrete 
ordinate methods, where the Boltzmann equation is directly discretized in
momentum space as well as in physical space
\citep[e.g.][]{Yueh77,Mezzacappa99, Liebendorfer04}, and moment-based
approaches, where angular integrations of the Boltzmann equation in
momentum space are performed \citep[e.g.][]{Thorne81,Burrows00,
  Rampp02}.  These two approaches give similar results in
one-dimensional models, at least in the context of core-collapse
supernovae \citep{Liebendorfer05}.  An additional technique that has
only been employed for solving static problems in the supernova
context, but is perhaps the most capable of retaining fidelity to the
underlying Boltzmann equation, is Monte Carlo neutrino transport
\citep{Janka89,Keil03}.

The moment approach results in an infinite hierarchy of coupled
equations which needs to be truncated at some order in practice.
Generally, only the zeroth and first order moment equations are
retained and a closure relation is assumed between the first two
moments and the higher order moments that enter the first two moment
equations.  Such schemes are referred to as variable Eddington factor
methods \citep{Mihalas84}.  When only the first two moments are used,
the number of equations relative to discrete ordinate methods is
significantly reduced, easing the computational burden (especially in
an implicit scheme like the one described below).  Of course, this
gain in computational efficiency is useful only if reasonable closures
can be obtained.  The closure relations only encode information about
the angular distribution of neutrinos, so that the approximations
involved in solving a linearized Boltzmann equation do not severely
impact the fidelity of numerical calculations to the true solution
\citep{Mihalas84,Ensman94}.
  
Here a variable Eddington factor approach to radiative transfer is
employed, with the closure relations being obtained from a formal
solution of the static relativistic Boltzmann equation.  This approach
is similar to that used by \citet{Burrows00} and \citet{Rampp02},
except for being fully general relativistic, incorporating both inelastic
scattering and pair production (in contrast to only the former), using
energy integrated groups rather than energy ``pickets'', and in the
specific method of finding the closure relations.  The formalism for
this method is described below.

The moments of the Boltzmann equation also most naturally give the various 
forms of the diffusion approximation, which has been used in the majority 
of PNS studies \citep{Burrows86,Keil95a,Pons99,Roberts12} and in a 
significant fraction of studies of the early core-collapse and bounce 
phases\citep{Bruenn85,Wilson93}.  The formalism is connected to EFLD 
in appendix \ref{sec:EFLD}.

\subsection{General Relativistic Generalities}

In spherical symmetry, it is simplest to work in a coordinate system
that anticipates a Lagrangian frame for the fluid.  The metric for
such a space-time is given by \citep{Misner64}
\be
ds^2 = -e^{2 \phi} dt^2 + \left(\frac{r'}
{\Gamma}\right)^2 da^2 + r^2 d \Omega^2,
\ee 
where $ds$ is the invariant interval, $t$ is the time measured
at infinity, $r$ is the areal radius, $\Omega$ is the solid angle,
and $\Gamma$ and $\phi$ are metric potentials.  Coordinate freedom 
can be exploited to choose this frame to be the rest-frame of 
the fluid, which demands \citep{Liebendorfer01}
\be
\label{eq:r_constraint}
\pd{r}{a} = \frac{\Gamma}{4 \pi r^2 n_B}.
\ee
Here, $n_B$ is the baryon number density and 
\be
\Gamma = \sqrt{1 + u^2 - \frac{2 m}{r}}
\ee
where $u$ and $m$ are defined below.
With this choice, the orthonormal frame associated 
with the coordinate frame is just the rest frame of the 
fluid and $da$ is just the change in enclosed baryon number
with the physical volume.  Therefore, this formulation is 
working in the Lagrangian frame, as claimed. 

The equations of spherically symmetric general relativistic 
hydrodynamics and the Einstein equation are recorded for 
convenience \citep{Misner64}.  Most of these results are 
nicely presented and detailed in similar form by
\cite{Liebendorfer01}.  The time evolution of the areal 
radius is given by, 
\be
\label{eq:r_evolve}
\pd{r}{t} = e^\phi u
\ee
which defines $u$.  The evolution of $u$ is given by  
\be
\label{eq:u_evolve}
\pd{u}{t} = \Gamma^2 \pd{e^\phi}{r} 
- e^\phi \frac{m + 4 \pi r^3 (p + Q)}{r^2},
\ee
where $Q$ is the viscosity and $p$ is the pressure of the fluid.
This gives the equation of hydrostatic balance when the 
left hand side equals zero (i.e. the Tolman-Oppenheimer-Volkov
equation \citep{Oppenheimer39}).  The enclosed gravitational 
mass, $m$, is defined by
\be
\label{eq:m_constraint}
\pd{m}{a} = \Gamma \left(\frac{E}{n_B}+ \epsilon \right) + 
u \frac{H}{n_B},
\ee
where $\epsilon$ is the internal energy per baryon, $E$ is 
the total neutrino energy density in the rest frame, and  
$H$ is the net radial energy flux from neutrinos.
The constraint equation for the metric potential $\phi$ 
is 
\be
\label{eq:phi_constraint}
\frac{\epsilon}{e^\phi} \pd{e^\phi}{a} 
+ \frac{1}{n_B e^\phi} \pd{(e^\phi p)}{a}
+ \frac{1}{r^3 n_B e^\phi} \pd{(r^3 e^\phi Q)}{a} = 0,
\ee
where a small time dependent term has been neglected.

The transport equations described in the next section 
are formulated in a congruence corresponding to the 
four-velocity field of the PNS (clearly, this is not 
a geodesic congruence).  The behavior of this 
congruence is best described by expanding the covariant
derivative of the four-velocity as
\be
U_{\mu;\nu} = -a_\nu U_\mu + \frac{\Theta}{3} P_{\mu \nu}
+ \sigma_{\mu \nu} + \omega_{\mu \nu},
\ee
where $U_\mu$ is the tangent four-vector field of the 
congruence, $P_{\mu \nu}$ is the projection tensor (which
projects into the vector subspace orthogonal to $U^\mu$)
$a^\nu = U^\alpha U^\nu_{;\alpha}$ is the acceleration,
$\Theta=U^\mu_{;\mu}$ is the expansion, $\sigma_{\mu \nu}$
is the shear, and $\omega_{\mu \nu}$ is the rotation.  Using 
the continuity equation, the expansion of the congruence 
becomes
\be
\Theta = - D_{\hat t} \ln(n_B).
\ee
In spherical symmetry, the acceleration four-vector is 
parallel to the radial orthonormal basis vector, so that
only the scalar acceleration is needed
\be
a = \Gamma \pd{\phi}{r}.
\ee
In spherical symmetry, the shear is characterized 
by a single component, the scalar shear
\be
\sigma = -\frac{2 u}{r} - \frac{2}{3} \Theta.
\ee
Additionally, such a spherically symmetric congruence
possesses no rotation, so that $\omega_{\mu\nu} = 0$. 
The quantity
\be
b = \frac{\Gamma}{r}
\ee
will also be required, which is related to the extrinsic 
curvature \citep{Thorne81}.  The orthonormal frame 
temporal and radial derivative operators are
\be
D_{\hat t} = e^{-\phi} \pd{}{t}
\ee
and
\be
D_{\hat r} = 4 \pi r^2 n_B \pd{}{a} = \Gamma \pd{}{r}.
\ee

\subsection{Variable Eddington Factor Transport Equations}
\label{Sec:MomentTransport} 

Here, the evolution equations for the neutrino number density, energy 
density, number flux and energy flux are derived from the zeroth and first 
order moments of the relativistic Boltzmann equation.  The basic results 
are taken from the spherically symmetric version of the projected symmetric 
trace-free moment formalism of \cite{Thorne81}.  This formalism reduces to 
an expansion of the neutrino distribution function in terms of Legendre 
polynomials in a flat space-time. 

The moments of the distribution function are defined in spherical symmetry as
\be
w^n = \frac{\omega^3}{(2 \pi)^2} B_n \int_{-1}^1 d \mu  P_n(\mu) f(\omega,\mu)
\ee  
where
\be
B_n = \frac{n! (2n + 1)}{(2n+1)!!},
\ee
$P_n$ are the Legendre polynomials, and $\omega$ is the neutrino energy in 
the fluids rest frame.  The first two moment equations 
in spherical symmetry can be read off from equation 5.10 of \cite{Thorne81}
\bea
w^0_{,\hat t} + \frac{4}{3} \Theta w^0
+ \frac{3}{2} \sigma w^2 + w^1_{,\hat r} &&\nonumber\\
2(a + b) w^1 - \pd{}{\omega} \omega
\left[a w^1 + \frac{\Theta}{3}w^0 + \frac{3}{2} \sigma w^2 \right]
&=& s^0 
\eea
and
\bea
w^2_{,\hat r} + (a + 3b)w^2 + w^1_{,\hat t} + \left[\frac{4}{3} \Theta 
+ \sigma \right]w^1 + \frac{1}{3} w^0_{,\hat r} + \frac{4}{3} a w^0 
\nonumber\\
- \pd{}{\omega} \omega \left[a w^2 + \left(\frac{\Theta}{3} + 
\frac{2}{5} \sigma\right)w^1 + \frac{1}{3}a w^0 + \frac{3}{2}\sigma 
w^3 \right] = s^1,
\eea
where $s^l$ are the neutrino source terms defined in section
\ref{sec:SourceTerms}.  To close this system, define the Eddington 
like factors
\be
g_2 = w^2/w^0
\ee
\be
g_3 = w^3/w^1
\ee
which both go to zero in the limit $f(\mu) = f_0 + \mu f_1$, which 
corresponds to the diffusion regime.  Note that these differ from the 
standard definition of the Eddington factors \citep{Rampp02}, which is 
due to how I have chosen to calculate the moments.  For free streaming 
radiation in a flat background $g_2 = 2/3$, and these equations reduce 
to the linear wave equation for $h=h(r \pm t) \equiv r^2 w_0$.  A method 
for approximating these Eddington factors is detailed in section 
\ref{sec:FormalSolution}.

For problems that are close to being static on the radiation timescale,
it is useful to switch the independent variable $\omega$, the energy in 
the fluid rest frame, of $w^i$ to the energy at infinity, $\nu$ 
\citep[c.f.][]{Schinder89}.  In the case of PNS cooling, the energy at 
infinity is much closer to being a constant of the motion and therefore 
a more natural variable.  Additionally, this choice simplifies the formal 
solution of the Boltzmann equation.  The moments of the distribution 
function are then
\be
w^i = w^i(r,\nu(\omega,r,t)) 
\ee
where the energy at infinity is defined as $\nu = e^{\phi(r,t)}\omega$.
This means that the replacement 
\be
\pd{w^i}{x} \rightarrow \pd{w^i}{x} + \pd{\nu}{x} \pd{w^i}{\nu}
\ee
needs to be made for all radial and time derivatives, resulting in
\bea
\label{eq:Thorne_moment_0}
\pd{w^0/n_B}{t} + \frac{w^0}{n_B} e^{\phi}\left( \frac{\Theta}{3} 
+ g_2 \frac{3}{2} \sigma \right)
+ \pd{}{a} \left(4 \pi r^2 e^{\phi} w^1 \right) \nonumber\\
- \frac{e^{\phi}}{n_B}\pd{}{\nu} \nu \left[ 
\left( \frac{\Theta}{3} + g_2 \frac{3}{2} 
\sigma \right) w^0 \right] + \frac{\nu}{n} \pd{\phi}{t} \pd{w^0}{\nu}
= e^\phi \frac{s^0}{n_B}
\eea 
and
\bea
\label{eq:Thorne_moment_1}
e^{-\phi} \pd{w^1}{t} + \left[\frac{4}{3} \Theta + \sigma \right]w^1
+ n_B e^{-\phi} 
\pd{}{a}\left[4 \pi r^2 e^{\phi} \left(\frac{1}{3} + g_2\right) w^0\right]
\nonumber\\
+ \left(\frac{2}{3} - g_2 \right)\left( a - b \right)w^0 
- \pd{}{\nu} \nu \left[ \left(\frac{\Theta}{3} + 
\frac{2}{5} \sigma + \frac{3}{2}\sigma g_3\right)w^1 
\right] 
\nonumber\\
+ e^{-\phi} \pd{\phi}{t} \nu \pd{w^1}{\nu}  = s^1.\nonumber\\
\eea

To easily deal with optically thick regions where the distribution
function may possess a sharp Fermi surface, energy integrated groups 
are used rather than discrete energy ``pickets''.  The group numbers, 
energies, number fluxes, energy fluxes, and source terms in group $g$ 
are defined by
\bea
N_g   = \int_{\omega_{g,L}}^{\omega_{g,U}}\frac{d\omega}{\omega} w^0, &&
F_g   = \int_{\omega_{g,L}}^{\omega_{g,U}}\frac{d\omega}{\omega} w^1,\nonumber\\
S^0_g = \int_{\omega_{g,L}}^{\omega_{g,U}}\frac{d\omega}{\omega} s^0, &&
S^1_g = \int_{\omega_{g,L}}^{\omega_{g,U}}\frac{d\omega}{\omega} s^1,\nonumber\\
E_g   = \int_{\omega_{g,L}}^{\omega_{g,U}}d\omega w^0, &&
H_g   = \int_{\omega_{g,L}}^{\omega_{g,U}}d\omega w^1,\nonumber\\
Q^0_g = \int_{\omega_{g,L}}^{\omega_{g,U}}d\omega s^0, && {\rm and} \, \,
Q^1_g = \int_{\omega_{g,L}}^{\omega_{g,U}}d\omega s^1.
\eea
Here, $\omega_{g,L}$ is the lower energy bound of an energy group
and $\omega_{g,U}$ is the upper bound.  Integrating over energy at 
infinity within groups gives
\be
N_g = \int_{\nu_{L,g}}^{\nu_{U,g}} \frac{d\nu}{\nu} w^0 
\, {\rm and} \,
E_g = e^{-\phi} \int_{\nu_{L,g}}^{\nu_{U,g}} d\nu w^0 
\ee
and similar expressions for $F_g$, $H_g$, and the source terms.  The operators 
$\int d\nu/\nu$ and $\int d\nu$ can then be applied to the ``red shifted'' 
equations.  The evolution the neutrino group number densities are described by 
\bea
\label{eq:Ng_Evo}
\pd{}{t} \left( \frac{N_g}{n_B} \right)
+ \pd{}{a} \left(4 \pi r^2 e^{\phi} F_g \right) &&
\nonumber\\
- \frac{e^{\phi}}{n_B}\left( \frac{\Theta}{3} + g_2 \frac{3}{2} \sigma 
- e^{-\phi} \pd{\phi}{t} \right) w^0 \biggr|^{\nu_U}_{\nu_L} 
= e^{\phi} \frac{S^0_g}{n_B}.  
\eea
The last term on the left hand side describes the red or blue 
shifting of neutrinos to other groups via compression and time
variation of the metric potential $\phi$.  If the group comprises 
energies from zero to infinity, the red shifting terms drop out 
and one is left with the standard number transport equation given 
in \cite{Pons99}.  Applying the number operator to 
equation \ref{eq:Thorne_moment_1} and simplifying gives 
\bea
\label{eq:Fg_Evo}
e^{-\phi} \pd{F_g}{t} + \left[\Theta + \left(\frac{3}{5} 
- \frac{3}{2}g_2 \right) \sigma \right] F_g
&&\nonumber\\
+ \frac{r^2 n_B}{3 e^{3\phi}} \pd{}{a} \left(4\pi e^{3\phi} N_g\right)
+ \frac{n_B}{r} \pd{}{a} \left(4 \pi r^3 g_2 E_g\right) 
&&\nonumber\\
- \left[\frac{\Theta}{3} + \left(\frac{2}{5} 
+ \frac{3}{2}g_2 \right) \sigma - e^{-\phi} \pd{\phi}{t}\right]
w^1\biggr|^{\nu_U}_{\nu_L} &=& S^1_g.
\eea
This includes similar terms to equation \ref{eq:Ng_Evo}, plus 
a term that includes the effects of compression on the
total number flux.  The energy group evolution equations are 
\bea
\label{eq:Eg_Evo}
\pd{}{t} \left( \frac{E_g}{n_B} \right)
+ e^\phi \left( \frac{\Theta}{3} + g_2 \frac{3}{2} \sigma \right) \frac{E_g}{n_B}
+ e^{-\phi} \pd{}{a} \left(4 \pi r^2 e^{2\phi} H_g \right) &&
\nonumber\\
- \frac{1}{n_B}\left( \frac{\Theta}{3} + g_2 \frac{3}{2} \sigma 
- e^{-\phi} \pd{\phi}{t} \right) \left(\nu w^0\right) \biggr|^{\nu_U}_{\nu_L} 
= e^{\phi} \frac{Q^0_g}{n_B}.  \nonumber\\
\eea
Aside from the addition of a compression term and different factors
of $e^\phi$, this is identical to equation \ref{eq:Ng_Evo}.  The energy 
flux group evolution equations are
\bea
\label{eq:Hg_Evo}
e^{-\phi} \pd{H_g}{t} + \left[\frac{4}{3} \Theta + \sigma \right] H_g
&&\nonumber\\
+ \frac{r^2 n_B}{3 e^{4 \phi}} \pd{}{a} \left(4\pi e^{4\phi} E_g\right)
+ \frac{n_B}{r e^\phi} \pd{}{a} \left(4 \pi r^3 e^\phi g_2 E_g\right) 
&&\nonumber\\
- \left(\frac{\Theta}{3} + \frac{2}{5} \sigma 
+ g_3 \frac{3}{2} \sigma  - e^{-\phi} \pd{\phi}{t}\right)
\left(\nu w^1\right)\biggr|^{\nu_U}_{\nu_L} &=& Q^1_g.
\eea
The numerical implementation of the red-shifting terms is described
in section \ref{sec:RedShift}.

Additionally, neutrinos have a back-reaction on the matter they 
are propagating through by exchanging energy, lepton number, and
momentum with the background medium.  Assuming that the background
possesses a thermal state, the first law of thermodynamics for the 
medium can be combined with the sum of equations \ref{eq:Eg_Evo} 
over all groups to find an equation for the conservation of total
internal energy
\bea
\label{eq:InternalEnergy}
\pd{}{t}\left(\epsilon + \sum_{g,s} \frac{E_{g,s}}{n} \right) 
+ e^{\phi}\Theta \left(\frac{p}{n} + \sum_{g,s} \frac{E_{g,s}}{3n} \right)
&&\nonumber\\
+  \frac{3e^{\phi}}{2}\sigma \sum_{g,s} g_{2,g} \frac{E_{g,s}}{n} 
+ e^{-\phi}\pd{}{a} \left(4 \pi r^2 e^{2 \phi} \sum_{g,s} H_{g,s} \right)
&=& 0
\eea
where the sums are over groups and species.  Obviously, the 
neutrino energy source terms have exactly canceled with the 
source terms for the medium.  

In the absence of neutrinos,
the electron fraction of the background medium is fixed, i.e.
$e^{-\phi}\dot Y_e = 0$.  When neutrinos are included, interactions
of electron flavored neutrinos exchange lepton number with the 
background, yielding $e^{-\phi}\dot Y_e = -\sum_g S^0_g/n_B$.  
The total lepton number of the medium is given by $Y_L = Y_e + \sum_{g} 
\left[N_{g,\nu_e} - N_{g,\bar \nu_e} \right]/n_B$.  Combining 
the evolution equation for $Y_e$ with equations \ref{eq:Ng_Evo} 
gives the lepton number evolution equation
\bea
\label{eq:LeptonFraction}
\pd{}{t}\left(Y_e + \sum_{g}
\left[ \frac{N_{g,\nu_e}}{n_B} - \frac{N_{g,\bar \nu_e}}{n_B} \right] \right) 
&& \nonumber\\
+ \pd{}{a} \left(4 \pi r^2 e^{\phi} \sum_g
 \left[F_{g,\nu_e} - 
F_{g,\bar \nu_e}\right]  \right) &=& 0.
\eea
This constitutes the full set of evolution equations for the 
state of the medium including non-thermal neutrinos of all 
flavors, when the Eddington factors $g_2$ and $g_3$ are 
specified.

\subsection{Neutrino Source Terms}
\label{sec:SourceTerms}   

The collision term in equation \ref{eq:Boltzmann} describes how
neutrinos move from one trajectory to another via scattering and 
how they are created and destroyed by the underlying medium.  For 
the PNS problem these processes include neutral current scattering 
off of electrons, nucleons, and nuclei \citep{Reddy98}, neutrino 
pair production via nucleon-nucleon bremsstrahlung \citep{Hannestad98}
and electron-positron annihilation \citep{Bruenn85}, and charged 
current processes involving electron and anti-electron flavor 
neutrinos and neutrons and protons, respectively \citep{Reddy98}.  

The details of these microphysical processes are eschewed by 
assuming that the differential cross-sections for these processes 
are known and referring the reader to the papers cited above, as 
well as the review \cite{Burrows06}. The exact details of the 
microphysics used in the code will be reported in a future publication,
although certain aspects are discussed in sections \ref{sec:PNSEvolution}
and \ref{sec:Spectra Compare}.  Many of the results in this section are 
well known \citep[e.g.][]{Bruenn85,Pons99}, and are included here for 
completeness and to make clear the details of the exact implementation
within the integrated energy group formalism described above.  Explicit 
detailed balancing (independent of the choice of underlying scattering kernels) 
is emphasized.
  
The source function for a particular moment is given by
\bea
&s^l = \frac{\omega^3}{(2 \pi)^2} B_l \int_{-1}^1 d\mu \, P_l(\mu)
&\nonumber\\
&\times \left(j_a(1-f) - \frac{f}{\lambda_a} + j_s(1-f) - \frac{f}{\lambda_s} 
+ j_p (1-f) - \frac{f}{\lambda_p}\right),&
\eea
which includes contributions from absorption $(1/\lambda_a)$, scattering 
$(1/\lambda_s)$, pair-annihilation $(1/\lambda_p)$, and their inverses 
$(j_a,j_s, \, {\rm and} \, j_p)$.  The choice of metric and reference frame 
implies that the scattering kernels should be evaluated in the rest frame 
of the fluid, simplifying things compared to hybrid frame approaches 
\citep{Hubeny07}.

For the absorption part, using the standard detailed balance relations gives
\be
j_a(1-f) - \frac{f}{\lambda_a} = 
\frac{1}{\lambda_a^*}\left(f_{eq}(\omega,T,\mu_{{\rm eq}})-f(\omega,\mu) \right), 
\ee
where $\lambda_a^{*-1} = [1+\exp\{-(\omega-\mu_{{\rm eq}})/T\})]\lambda_a^{-1}$
and $f_{eq}$ is a Fermi-Dirac distribution.
The scattering contributions are 
\be
j_s = \int \frac{d\omega'}{(2\pi)^3} \omega'^2 \int^{1}_{-1} d\mu' 
\int_0^{2\pi} d\phi' \, R^s(\omega',\omega,\mu')f(\omega',\mu_{out})
\ee
and 
\be
\lambda_s = \int \frac{d\omega'}{(2\pi)^3} \omega'^2 \int^{1}_{-1} d\mu' 
\int_0^{2\pi} d\phi' \, R^s(\omega,\omega',\mu') (1-f(\omega',\mu_{out})),
\ee
and the pair-annihilation contributions are
\be
j_p = \int \frac{d\omega'}{(2\pi)^3} \omega'^2 \int^{1}_{-1} d\mu' 
\int_0^{2\pi} d\phi' \, R^p_{in}(\omega,\omega',\mu')(1-\bar f(\omega',\mu_{out}))
\ee
and
\be
\lambda_p = \int \frac{d\omega'}{(2\pi)^3} \omega'^2 \int^{1}_{-1} d\mu' 
\int_0^{2\pi} d\phi' \, R^p_{out}(\omega,\omega',\mu') \bar f(\omega',\mu_{out}).
\ee
The outgoing cosine is given by $\mu_{out} = \mu\mu'-\sqrt{1-\mu^2}
\sqrt{1-\mu'^2}\cos \phi'$.  The $R_{out}$ functions are related to
the differential cross-section by
\be
R(\omega,\omega',\mu) = \frac{(2 \pi)^2}{\omega'^2}\frac{1}{V}
\frac{d \sigma}{d \omega' d\mu},
\ee
with no phase space blocking term for the final neutrinos in the 
differential cross-section.  The $R$ functions obey the detailed 
balance relations for scattering
\be
R^{s}(\omega,\omega',\mu) = R^{s}(\omega',\omega,\mu)e^{(\omega-\omega')/T}.
\ee
and annihilation
\be
R^p_{out}(\omega,\omega',\mu) = R^p_{in}(\omega,\omega',\mu) 
e^{(\omega + \omega')/T}
\equiv R^p(\omega,\omega',\mu).
\ee
For use in the moment formalism, $R$ must be expanded in terms of 
the Legendre polynomials as 
\be
R(\omega,\omega',\mu') = \sum_{l=0}^\infty R_l(\omega,\omega') P_l(\mu').
\ee
The distribution function, $f$, is also expanded in a similar way.
In general, this results in integrals of the form
\be
\ba{rl}
F_{klmn} =& \int_{-1}^{1}d\mu \int_{-1}^1d\mu' \int^{2 \pi}_0 d \phi'
P_k(\mu') P_l(\mu) P_m(\mu) \\
&\times P_n(\mu \mu' - \sqrt{1-\mu^2}\sqrt{1-\mu'^2}\cos\phi')\\
=& 2 \pi \frac{\delta_{kn}}{2n + 1} I_{lmn}. 
\ea
\ee
where $I_{lmn} = \int_{-1}^{1} d\mu P_l(\mu) P_m(\mu) P_n(\mu)$. For 
a more detailed description of such an expansion, see \cite{Mezzacappa93}.

Using this expansion, the scattering contribution to the source term is 
given by
\bea
\label{eq:sl_scatter}
s^l_s =&& \frac{4 \omega^3}{(2 \pi)^4} B_l \int d\omega' \omega'^2
\nonumber\\
&& \times \Biggl\{
\frac{R_l^s f_l'}{(2 l + 1)^2}e^{-(\omega-\omega')/T}  - 
\frac{R_0^s f_l}{2l + 1}
\nonumber\\
&&+\frac{1}{2} \sum_{m,n=0}^\infty R_n^s f_n'f_m \frac{I_{lmn}}{2n+1}
\left(1 - e^{-(\omega-\omega')/T}\right)
\Biggr\},
\eea
and the pair annihilation contribution is given by
\bea
\label{eq:sl_pair}
s^l_p =&& \frac{4 \omega^3}{(2 \pi)^4} B_l \int d\omega' \omega'^2
e^{-(\omega+\omega')/T}
\nonumber\\
&& \times \Biggl\{
R^p_0 \delta_{0l} -  \frac{R^p_0 f_l}{2l + 1} - \frac{R^p_l \bar f_l'}{(2l + 1)^2}
\nonumber\\
&&+\frac{1}{2} \sum_{m,n=0}^\infty R_n^p f_m \bar f_n'  \frac{I_{lmn}}{2n+1}
\left(1 - e^{(\omega+\omega')/T}\right)
\Biggr\}.
\eea

Clearly, all of the moments are coupled to all of the other moments by
the source terms in addition to the coupling present on the LHS of the
moment equations.  Practically, this series must be truncated at some
finite order.  It is standard to use only the zeroth and first moment
\citep{Burrows06}.  This convention is followed, but with the caveat
that this may not be a good approximation for the annihilation terms
near the free streaming regime \citep{Pons98}.

\subsubsection{Zeroth Order Source Function}

Now the three contributions to the source functions for the number 
and energy group equations are considered separately.  
Special attention is given to assuring that the chosen forms 
for the source terms {\it explicitly} push the neutrinos towards 
equilibrium, independent of the chosen opacity functions.

The absorption part of the source function is given by
\be
S^0_{a,g} = \left\langle \frac{1}{\lambda_a^*} \right\rangle_g 
\left[G_g - N_g \right],
\ee
and 
\be
Q^0_{a,g} = \left\langle \frac{1}{\lambda_a^*} \right\rangle_g 
\left[B_g - E_g \right],
\ee
where 
\be
B_g = \int_{\omega_L^g}^{\omega_U^g} d \omega \frac{2 \omega^3}{(2 \pi)^2} 
\frac{1}{e^{(\omega-\mu_{{\rm eq}})/T} + 1},
\ee 
and 
\be
G_g = \int_{\omega_L^g}^{\omega_U^g} d \omega 
\frac{2 \omega^2}{(2 \pi)^2} \frac{1}{e^{(\omega-\mu_{{\rm eq}})/T} + 1}.
\ee 
The average over the inverse absorption mean free path
can be performed in a number of ways.  For small enough
energy intervals for groups, the mean free path for the 
central energy of the group can be taken.  It can also be 
assumed that the energy within a group is distributed as 
in a blackbody, but renormalized to the total energy within 
the group.  Then the averaged inverse mean free path is analogous 
to the Planck mean opacity, but using a Fermi-Dirac distribution
instead of a Planck distribution.  Independent of the 
averaging procedure chosen, this term serves to push the
neutrino energy density towards equilibrium with the 
background medium.

To find the  scattering contribution to the zero order moment equation,
it is assumed that the distribution of energy within a particular energy 
group is proportional to the blackbody distribution.  Using this ansatz 
in equation \ref{eq:sl_scatter} and then averaging over group energies 
gives scattering term
\bea
S^0_{s,g} = \sum_{g'} \Phi_{0,gg'}^s  
N_{g'} \left(\frac{D_{g'}}{G_{g'}} - 1\right)
\left(D_g - N_g\right)
\nonumber\\
- \Phi_{0,gg'}^s 
N_g \left(\frac{D_{g}}{G_{g}} - 1\right)\left(D_{g'} 
- N_{g'} \right) 
\nonumber\\
+ \Phi_{1,gg'}^{s}F_g F_{g'}
\left(\frac{D_{g}}{G_{g}}  - \frac{D_{g'}}{G_{g'}}\right),
\eea
and
\bea
Q^0_{s,g} = \sum_{g'} \Phi_{0,gg'}^s  
N_{g'} \left(\frac{D_{g'}}{G_{g'}} - 1\right)
\left(C_g - E_g\right)
\nonumber\\
- \Phi_{0,gg'}^s 
E_g \left(\frac{C_{g}}{B_{g}} - 1\right)\left(D_{g'} 
- N_{g'} \right) 
\nonumber\\
+ \Phi_{1,gg'}^{s}H_g F_{g'}
\left(\frac{C_{g}}{B_{g}}  - \frac{D_{g'}}{G_{g'}}\right),
\eea
where 
\be
C_g = \int_{\omega_L^g}^{\omega_U^g} d \omega \frac{2 \omega^3}{(2 \pi)^2}, \, {\rm and} \,
D_g = \int_{\omega_L^g}^{\omega_U^g} d \omega \frac{2 \omega^2}{(2 \pi)^2}.
\nonumber
\ee
The averaged scattering kernel is defined as  
\be
\Phi_{l,gg'}^s = \left \langle R_l(\omega,\omega')
e^{-(\omega-\mu_{\nu,\rm{eq}})/T} \right \rangle_{\Delta E_g,\Delta E_{g'}},
\ee
where the average is taken over the energies of the incoming
and outgoing groups.  
This has the useful property $\Phi_{l,gg'}^s = \Phi_{l,g'g}^s$, which
derives from the detailed balance criterion given above.  This form
of the scattering source term naturally conserves neutrino number, 
although it does not push the neutrino numbers toward the expected 
distribution for a purely scattering process.  Both source terms go to 
zero when neutrinos are in both chemical and energy equilibrium with the
medium.  The energy exchange expression does not go to zero when $g'=g$.  
Although it might be naively assumed that this corresponds to elastic 
scattering and therefore not contribute to the evolution of the group 
energy, there are in fact contributions from
small energy transfer scatterings to this term as well.  Although these
will conserve neutrino number within the group, they can result in 
energy exchange with the medium.  This allows the formalism to somewhat 
naturally deal with energy transfer due to scattering off nucleons, which 
generally exchanges energy on a scale that is smaller than the group 
spacing.  The weighting function for the average over the groups 
necessarily involves some level of approximation.  The natural 
incorporation of equilibrium far outweighs the small error introduced 
due to the approximate weighting of the scattering kernel.
	
Using a similar procedure to the one used for the scattering source
term, the energy integrated pair production/annihilation source term
is given by
\bea
S^0_{p,g} = \sum_{g'} 
\Phi_{0,gg'}^p 
\left(D^0_{g'} - \bar N_{g'} \right) \left(D^0_{g} - N_{g} \right) 
\nonumber\\
- \Phi_{0,gg'}^p N_g \bar N_{g'}
\left(D^0_{g'}/\bar G_{g'} - 1 \right) \left(D^0_{g}/G_{g} - 1\right) 
\nonumber\\
+ \Phi_{1,gg'}^p F_g \bar F_{g'}
\left(1 - e^{(\omega+\omega')/T}\right),
\eea
where the over bar denotes the energy density and flux of a neutrinos 
anti-species.  This term once again naturally goes to zero when thermal 
equilibrium is reached (i.e. when $N_g = G_g$ and $F_g = 0$).  Here,
\be
\Phi_{i,gg'}^p = \left \langle e^{-(\omega+\omega')/T}
R_i^p(\omega,\omega') \right \rangle_{\Delta E_g,\Delta E_{g'}}.
\ee
The source term $Q^1_p$ can be obtained from the above equation
by the replacements $F_g \rightarrow H_g$, $N_g \rightarrow E_g$,
$D_g \rightarrow C_g$, and $G_g \rightarrow B_g$, while leaving 
the $g'$ terms unchanged.

\subsubsection{First Order Source Function}

For the first order source function terms, one does not need to 
be as careful about getting forms that explicitly go to zero in 
equilibrium as all terms end up being proportional to the first 
order distribution function and therefore satisfy this constraint
automatically.

The absorption contribution to the first order source function is
\be
S^1_a = -  F_g \left\langle \frac{1}{\lambda_a^*} \right\rangle_g,
\ee
and 
\be
Q^1_a = -  H_g \left\langle \frac{1}{\lambda_a^*} \right\rangle_g.
\ee

The scattering source term in equation \ref{eq:Ng_Evo} is
\bea
S^1_s = \sum_{g'}
\Phi^s_{0,gg'} F_g
\left[ N_{g'} \left(\frac{D_g}{G_g} - \frac{D_{g'}}{G_{g'}}\right)
- D_{g'}\left(\frac{D_{g}}{G_{g}}-1\right) \right]
\nonumber\\
+\frac{\Phi^s_{1,gg'}}{3}F_{g'}  
\left[N_g \left(\frac{D_g}{G_{g}} - \frac{D_{g'}}{G_{g'}}\right) 
+ D_g\left(\frac{D_{g'}}{G_{g'}}-1\right)\right].
\eea
The source term $Q^1_s$ can be obtained from the above equation
by the replacements $F_g \rightarrow H_g$, $N_g \rightarrow E_g$,
$D_g \rightarrow C_g$, and $G_g \rightarrow B_g$, while leaving 
the $g'$ terms unchanged.  When scattering is iso-energetic, this 
reduces to
\bea
S^1_{s,{\rm iso-en}} = -F_g \frac{2 \omega^2}{(2 \pi)^2}
\left[\tilde R_0^s(\omega) - \tilde R_1^s(\omega)/3 \right]
\nonumber\\
\equiv -H_g \left[\chi_0^s(\omega) - \chi_1^s(\omega)/3 \right]
\eea
where $\tilde R_l^s$ is the iso-energetic scattering kernel.

The first order moment equation pair annihilation source term is
\bea
S^1_p = \sum_{g'}
\Phi^p_{0,gg'}  F_g
\left[\bar N_{g'} \left(1 - e^{\beta(\omega+\omega')}\right) - 
D_{g'}^0\right] \nonumber\\
+ \frac{\Phi^s_{1,gg'}}{3} \bar F_{g'} \left[ N_g   
\left(1 - e^{\beta(\omega+\omega')}\right) - D_g^0\right],
\eea
and $Q^1_p$ can be found the same replacement required to find
$S^1_s$ from $Q^1_s$.

\section{Formal Solution of the Boltzmann Equation}
\label{sec:FormalSolution}

To get the factors $g_2$ and $g_3$, an angle dependent version of the
Boltzmann equation needs to be solved. First, note that the outer
layers of the PNS are in tight radiative equilibrium throughout the
duration of the simulation.  Therefore, all time dependence can be
reasonably dropped in the equation of radiative transfer if one is
only interested in the ratios of various moments.  Of course, such an
approximation breaks down in highly dynamical situations.  For such
circumstances, a closure scheme like the one described in \cite{Rampp02}
is more appropriate.  This time-independent formulation of the formal
solution, which makes calculation of the Eddington factors
significantly easier, is similar to the approach advocated by
\cite{Ensman94}, except that it incorporates general relativistic
affects, such as the bending of geodesics.  \cite{Schinder89} describe
a similar, but time-dependent formulation.

In a spherically symmetric static spacetime, the equation of radiative 
transfer is \citep{Lindquist66}
\bea
\label{eq:boltz}
\Gamma \left[ \mu \pd{}{r} + (1-\mu^2) \left\{\frac{1}{r} - \pd{\phi}{r} \right\} 
\pd{}{\mu}\right] f(\nu,\mu,r) \nonumber\\
= \frac{1}{\lambda^*_a} \left[f_{eq}(r) - f(\nu,\mu,r)\right]
\nonumber\\
+ j_s[f](1-f(\nu,\mu,r)) - \lambda_s^{-1}[f]f(\nu,\mu,r).
\eea
Here, the neutrino distribution function, $f$, has been written in terms of the 
energy of the neutrinos at infinity, $\nu$, and $\mu$ is the cosine of the angle 
of neutrino propagation relative to the radial vector.

A formal solution to equation \ref{eq:boltz} can easily be found using the method 
of characteristics.  The characteristic equations are 
\be
d\lambda \equiv \frac{dr }{\Gamma \mu} = \frac{d\mu }
{\Gamma (1-\mu^2)(1/r - \pd{\phi}{r})}
= df\bigr/\left(\frac{df}{d\lambda} \right)_{coll}
\ee
where $\lambda$ is the physical path length.  The second equality is easily integrated
to find a relationship between $r$ and $\mu$ along a geodesic.  Any geodesic can be 
characterized by the radius at which $\mu = 0$.  First, define the quantity 
\be
\beta = r_m e^{-\phi_m},
\ee
where the subscript $m$ denotes the minimum radius of propagation.  This is just
the impact parameter of the trajectory.  Then, for a given $\beta$ and $r$, the 
angle of propagation along a geodesic is given by
\be
\mu = \pm \sqrt{1-\left(\frac{\beta e^{\phi}}{r}\right)^2}.
\ee

The first equality in the characteristic equations can be integrated to 
find the physical path length between any two radii for a particular 
characteristic if $\Gamma$ and $\phi$ are assumed constant over this 
distance, giving
\be
\label{eq:path_length}
\Delta \lambda \approx \pm \Gamma^{-1} \left[\sqrt{r_f^2-e^{2\phi}\beta^2}
-\sqrt{r_i^2-e^{2\phi}\beta^2}\right],
\ee
where the plus sign is for $r_f>r_i$ and the minus sign otherwise.  This 
form is consistent with the assumption of constant metric functions across
zones (as is used in the actual code), but it can introduce difficulties when 
a trajectory moves from one zone to another near the radius of minimum propagation.

Clearly, equation \ref{eq:boltz} is a non-linear integro-differential equation due 
to the functional dependence of the scattering terms on the local distribution 
function.  An approximate solution to the Lindquist equation is desired were the 
solutions along characteristics are decoupled and the formal solution can be directly
integrated.  The simplest approximation is to just to make the replacement 
$f \rightarrow f_{eq}$ in the scattering terms.  This approximation will only be 
valid at high optical depth and is therefore suspect for use in the decoupling 
region.  The next order approximation is to use a distribution function inferred 
from our knowledge of $E_g$ and $H_g$.  Assuming that the energy is distributed 
within a group as within a blackbody gives
\bea
\hat f_0(\nu,r) = f_{{\rm eq}}(\nu,r) \frac{E_g}{B_g},
\nonumber\\
\hat f_1(\nu,r) = 3 f_{{\rm eq}}(\nu,r) \frac{H_g}{B_g}.
\eea 
Employing the Legendre expansion of the scattering kernel, integrating over 
outgoing neutrino angle, and only including the elastic scattering contribution
gives the scattering source and sink terms
\bea
j_s(\omega,\mu) &=& \frac{2 \omega^2}{(2 \pi)^2} \sum_{l=0}^{\infty} \frac{1}{2l + 1}
P_l(\mu) \tilde R^s_l(\omega) \hat f_l(\omega) \nonumber\\
&\approx & f_{{\rm eq}}(\omega) \left\{\chi^s_0(\omega) \frac{E_g}{B_g}
+ \mu \chi^s_1(\omega) \frac{H_g}{B_g}  \right\}
\eea
and 
\be
\lambda_s^{-1}(\omega,\mu) = \chi^s_0(\omega)
- j_s(\omega,\mu).
\ee

Using the last characteristic equation, the solution of the 
linearized Boltzmann equation is  
\bea
\label{eq:formal_solution}
f(\nu,\beta,r_f) &&= f(\nu,\beta,r_i)e^{-\tau(r_i,r_f)}
\nonumber\\
&&+ e^{-\tau(r_i,r_f)} \int_{r_i}^{r_f} \frac{dr}{\Gamma \mu}
e^{\tau(r_i,r)}
\left\{j_s + f_{eq}/\lambda^*_a\right\},
\eea
where the optical depth is 
\be
\tau(r_i,r_f) = \int_{r_1}^{r_2} \frac{dr}{\Gamma \mu} 
(1/\lambda_a^* + \chi^s_0).
\ee
This has the appealing property that there is no coupling 
between different $\beta$s and $\nu$s, so the evolution of 
the distribution function along each path in phase space 
can be solved for independently.

\section{Numerical Implementation}

\label{sec:Numerics}
Aside from the equation of state and neutrino opacities for dense 
matter, PNS evolution is described by the transport equations 
\ref{eq:Ng_Evo}, \ref{eq:Fg_Evo}, \ref{eq:Eg_Evo}, 
\ref{eq:Hg_Evo}, \ref{eq:InternalEnergy}, and 
\ref{eq:LeptonFraction} and the structure equations 
\ref{eq:r_evolve}, \ref{eq:u_evolve}, \ref{eq:m_constraint}, 
\ref{eq:r_constraint}, and \ref{eq:phi_constraint}.  These 
describe the evolution of the dependent variables $y(a,t) = 
\left\{r,u,m,\phi,n_B,T,Y_e,F_g,N_g,E_g,H_g\right\}$.  To solve these
equations numerically, the variables $\{a_{i+1/2},r_{i+1/2},
u_{i+1/2},m_{i+1/2},F_{g,i+1/2},H_{g,i+1/2}\}$ are discretized on zone 
edges while the variables $\left\{\phi_i,n_{B,i},T_i,Y_{e,i},
N_{g,i},E_{g,i}\right\}$ are discretized on zone centers.  The derivatives
in the PNS evolution equations are then finite differenced, turning
them to algebraic equations for the above dependent variables.
This is the most natural choice for discretizing the above equations, 
because the thermodynamic quantities, neutrino number density, and 
neutrino energy density are defined on zone centers while the neutrino
number and energy fluxes are defined across zone edges which results in
internal energy and lepton number conservation being made explicit in 
the discretized equations.

The general form of these algebraic equations is then
\bea
\mathcal{G}\left(y_{i-1}^{n,n+1},y_{i}^{n,n+1},y_{i+1}^{n,n+1}\right) =
\mathcal{T}(y_{i}^n,y_{i}^{n+1}) 
\nonumber\\
+ (1-\theta) \, \mathcal{Y}\left(y_{i-1}^{n},y_{i}^{n},y_{i+1}^{n}\right)
+ \theta \,     \mathcal{Y}\left(y_{i-1}^{n+1},y_{i}^{n+1},y_{i+1}^{n+1}\right)
=0,
\eea 
where $n$ is the current time, at which the dependent variables 
are known, and $n+1$ is the next time step at which the dependent
variables are desired.  Here, $\mathcal{T}$ denotes the differenced time 
derivatives and $\mathcal{Y}$ denotes the rest of the terms.  I 
choose to employ a fully implicit method for solving these 
equations, i.e. $\theta = 1$.  This leaves a set of non-linear 
algebraic equations that must be solved to find the values of 
the dependent variables at time step $n+1$.

These equations are solved by standard high-dimensional Newton-Raphson
(NR) techniques \citep{Press92}.  This requires calculating
derivatives of all the functions $g$ with respect to $y$.  These
derivatives are calculated analytically.  Due to the number of
derivatives, such an undertaking is prone to error.  Therefore, all
derivative functions are checked against numerical derivatives by
automated software before they are included in the actual evolution
code.  The NR updates are given by the solution of an $N_z \times (6 +
2 N_g N_s)$ -by-$N_z \times (6 + 2 N_g N_s)$ matrix, where $N_z$ is
the number of radial zones, $N_g$ is the number of neutrino energy
groups, and $N_s$ is the number of included neutrino species.  This
can rapidly become quite large for reasonable zoning and number of
energy groups, and become too slow for dense matrix techniques.
Luckily, the matrix involved is in fact block-diagonal, as each zone
is only coupled to its neighboring zones, which significantly reduces
computational time compared to solving a general dense matrix.

Although the equations are formally non-linear, they are sufficiently
close to linear that NR iteration results in good convergence after a
small number of iterations.  It is generally demanded that the average
relative deviation of the solution from zero is at least less than one
part in a thousand.  Often, the solution found by NR iteration
satisfies the equations to close to machine precision.  This scheme
has been implemented using object-oriented \texttt{FORTRAN2003}.  The
block diagonal matrix equations are solved using the software package
\texttt{LAPACK} \citep{LAPACK}.

To save computational time, the equations are solved using only the
neutrino number equations and approximating the neutrino energy
densities and fluxes using $E_g \approx \langle \omega \rangle_g N_g$.
Once this set of equations is satisfied, a correction step is taken
using the energy groups instead of the number groups.  This
approximation does not seem to introduce any significant error into
the calculation.  It is found that the total neutrino energy loss
calculated using the approximation $E_g \approx \langle \omega \rangle_g 
N_g$ differs from the actual neutrino energy loss by around one part in
a thousand when thirty energy groups are used.  The code conserves lepton 
number to machine precision because lepton number conservation is explicitly
enforced by equation \ref{eq:LeptonFraction}.  Conservation of total energy
is not explicitly enforced.  It is found that the total change in rest mass 
over the simulation agrees with the total neutrino energy lost to within a 
few percent.  A series of test problems are performed with the
code in Appendix \ref{sec:CodeTests}.

\subsection{Equation of State}

To close the transport and structure equations described above, an
equation of state is required relating the pressure, energy density,
and equilibrium neutrino chemical potential to $n_B$, $T$, and $Y_e$.
Additionally, accurate derivatives of these quantities are required
for calculation of the Jacobian matrix for NR iteration.  Calls to the
equation of state must also be computationally efficient.  To meet
these requirements, the equation of state is implemented in a tabular
form.  The Helmholtz free energy per baryon, $F = \epsilon - sT$, is
tabulated as a function of $n_B$, $T$, and $Y_e$, as well as
derivatives with respect to these variables up to second order.  A
bi-quintic interpolation is then used to get values of the free energy
and its derivatives between grid points \citep{Timmes00}.  This
guarantees that the thermodynamic functions will be smooth in the
independent variables, thermodynamically consistent \citep{Swesty96},
and does not introduce problems in the calculation of the NR
corrections.

The differential of the Helmholtz free energy is 
\bea
dF &=& -s dT + \frac{p}{n_B^2} dn_B + \sum_i \mu_i d Y_i \nonumber\\
   &=& -s dT + \frac{p}{n_B^2} dn_B + (\mu_e + \mu_p - \mu_n) d Y_e.
\eea
From this, the required thermodynamic quantities can be read off:
\bea
p = n_B^2 \thd{F}{n_B}{T,Y_e}, \, s = -\thd{F}{T}{n_B,Y_e}, \nonumber\\
{\rm and} \, \mu_{\nu_e,{\rm eq}} \equiv (\mu_e + \mu_p - \mu_n) = 
\thd{F}{Y_e}{n_B,T}.
\eea

\subsection{Neutrino Opacities}

The group averaged neutrino opacities are calculated using a ten
point quadrature over each group to find an effective Planck
mean opacity for the absorption terms in each group.  The scattering 
and annihilation kernels which couple the groups, $\Phi_{g,g'}$, are
calculated using a five point quadrature over both the incoming and 
outgoing energies.  Detailed balance is exploited to halve the number 
of calculations required.  The scattering terms are not weighted by a 
local thermal neutrino distribution.

\subsection{Integration of the Formal Solution}

The formal solution to the static Boltzmann equation is calculated 
at the beginning of every time step and the Eddington factors 
enter the moment transport equations explicitly.  Because time 
independent transport is assumed, no previous knowledge of the 
distribution functions is required and a new grid of impact 
parameters can be chosen at any time step, without having to 
worry about re-mapping old solutions as in \cite{Rampp02}.  

If all quantities are assumed to be constant across zones, the 
formal solution (equation \ref{eq:formal_solution}) can easily 
be integrated, giving
\be
f(\nu,\mu_s,r_{i+1/2}) = f(\nu,\mu_s,r_{i-1/2}) e^{-\Delta \tau_i}
+ \Delta f_0 + \Delta f_1
\ee
for the change in $f$ across zone $i$.

The physical path length across the zone, $\Delta \lambda_i$, 
is given by  equation \ref{eq:path_length} and the optical 
depth across the zone is 
\be
\Delta \tau_i = \Delta \lambda_i 
(1/\lambda_a^* + \chi^s_0).
\ee
The additions to the neutrino beam from the medium and 
scattering from other beams are given by
\be
\Delta f_0 = f_{{\rm eq}}\frac{1/\lambda_a^* + 
\frac{E_g}{B_g}\chi^s_0}
{1/\lambda_a^* + \chi^s_0} (1-e^{-\Delta \tau_i}),
\ee
and
\be
\Delta f_1 = f_{{\rm eq}}\chi^s_1
\frac{H_g}{B_g} e^{-\Delta \tau_i}\int_0^{\Delta \lambda_i} d\lambda
 \mu(\lambda) 
e^{\lambda(1/\lambda_a^* + \chi^s_0)}.
\ee
The integral required for $\Delta f_1$ cannot be calculated 
analytically because $\mu(\lambda)$ is a fairly complicated function.
As this is a subdominant term, an ``average'' $\mu$ can be pulled out 
of the integral (which is allowable if $\mu$ does not change much across the 
zone).  This gives the approximation
\be
\Delta f_1 \approx \mu(r_i,\beta) f_{{\rm eq}}\chi^s_1\frac{H_g}{B_g} 
\frac{1-e^{\Delta \tau_i}}{1/\lambda_a^* + 1/\lambda_s^*}.
\ee
Note that $\mu$ changes most rapidly when it is close
to zero, but this term contributes the least in that region so the 
error from this approximation should not be too large.

Numerically, there is a problem with this formulation as it stands.
Assume that a trajectory in zone $i$ is close to its minimum radius 
of propagation, $\mu>0$, and that it is close to a zone boundary.  
It then propagates to the zone boundary and is considered to be in zone 
$i+1$.  Because $\phi$ is increasing with radius, $\phi_{i+1}>\phi_{i}$.  
The new radius is taken to be $r_{L,i+1}$, so that the new angle of 
propagation is 
\be
\mu_n = \sqrt{1-\left(\frac{r_m e^{\phi_{i+1}-\phi{i}}}{r_{L,i+1}} \right)^2},
\ee
$e^{\phi_{i+1}-\phi{i}}>1$, and $r_m \approx r_{L,i+1}$.  Since, $\mu_n$ 
must be real, it becomes ill defined.  In practice this problem is overcome 
setting $\mu$ to zero if it would have been imaginary.

Starting from the outer boundary of the computational grid, these 
equations are solved along an inward going characteristic, through
the radius of minimum propagation, and then along the outward going 
characteristic for each tangent ray.  The impact parameters of the 
tangent ray grid are chosen to be equally spaced in radius for the 
calculations described in this paper. Once the distribution function
for a particular energy at infinity has been calculated along tangent
rays, moments of the distribution function at radii $r_i$ are 
calculated from a weighted sum that reduces to the correct limit if 
the distribution function is locally constant in angle.  Angular resolution 
is reduced at larger depths in the star.  Because the distribution is 
extremely close to isotropy and $g_2 \approx g_3 \approx 0$, this does 
not pose a significant problem for PNS evolution.      

\subsection{Boundary Conditions}

To close the system of transport equations, boundary conditions 
for the surface fluxes $H_g$ and $F_g$ are required.  For this 
boundary condition, the formal solution is used to calculate the 
factors 
\be
\alpha_g = \frac{\int_{-1}^1 d\mu \mu f(r,\mu,\nu_g)}
{\int_{-1}^1 d\mu f(r,\mu,\nu_g)},
\ee 
so that $F_{g,{\rm bound}} = \alpha_g N_g$ and $H_{g,{\rm bound}} 
= \alpha_g E_g$ in the final zone.  At the inner edge of the 
computational grid, incident fluxes are specified (for PNS evolution,
they are of course specified to be zero).  

The boundary conditions for the radius, 
gravitational mass, velocity, and pressure are implemented by 
including a fixed ghost zone at the inner and outer boundaries.  
The boundary condition for the metric potential $\phi$ is given 
by matching to the Schwarzschild vacuum solution to the Einstein 
equations at the outer boundary.  This gives $\phi_s = \log(\Gamma_s)$.

\subsection{Rezoning}

To maintain reasonable spatial resolution, conservative post time step
re-gridding is employed.  Where conservation laws do not specify the 
properties of a new zone, piecewise linear interpolation is used.  This 
generally results in smooth radial dependence of the fluid quantities.
The implementation is similar to the method used in \texttt{Kepler} 
\citep{Weaver78}. 

The re-gridding is driven by gradients in the density and radius.  Generally,
the radius is not allowed to vary by more than $5\%$ between zones and 
the density is allowed to vary by no more than $20\%$.  This generally
results in approximately 100-150 zones being on the grid.  The choice 
of relative density changes places high resolution in regions where 
neutrino decoupling is occurring.

\subsection{Red Shifting Terms}
\label{sec:RedShift}

Due to red and blue shifting between groups, equations \ref{eq:Ng_Evo}, 
\ref{eq:Fg_Evo}, \ref{eq:Eg_Evo} and \ref{eq:Hg_Evo} contain the 
un-integrated moments $w^i$.  Therefore, an approximation method
for these moments is required.  When integrated 
over all energies, these terms go to zero.  Therefore, any chosen 
numerical scheme must have terms balancing between groups for energy 
conservation.  To move forward, something must be assumed about how 
energy is distributed in the groups.  The simplest scheme is to assume 
that it is uniform.  Then within a particular group $w^{0,1} = 
\{N_g,F_g,E_g,H_g\}/(\omega_{g,H}-\omega_{g,L})$.  It could also be 
assumed that the internal energy is distributed as a black body, which 
is consistent with the assumption used in the source terms. The uniform 
distribution is chosen due to its simplicity.  For the $H_g$ evolution 
equation, this results in
\be
\ba{c}
- \omega_{g,U} \biggl[ \left(\frac{\Theta}{3} + \frac{2}{5} \sigma 
+ g_3 \frac{3}{2} \sigma  - e^{-\phi} \pd{\phi}{t}\right)
\left( \frac{H_{g}}{2\Delta \omega_{g}} 
+ \frac{H_{g+1}}{2\Delta \omega_{g+1}}\right) \biggr]\\
+ \omega_{g,L} \biggl[ \left(\frac{\Theta}{3} + \frac{2}{5} \sigma 
+ g_3 \frac{3}{2} \sigma  - e^{-\phi} \pd{\phi}{t}\right)
\left( \frac{H_{g-1}}{2\Delta \omega_{g-1}} 
+ \frac{H_{g}}{2\Delta \omega_{g}}\right)\biggr].
\ea
\ee
Similar expressions result for equations \ref{eq:Ng_Evo}, 
\ref{eq:Fg_Evo}, and \ref{eq:Eg_Evo}.  It is straight forward to 
verify that these terms disappear when summed over groups.   

\section{Proto-Neutron Star Evolution}
\label{sec:PNSEvolution}

\begin{figure*}
\begin{center}
\leavevmode
\includegraphics[width=0.32\textwidth]{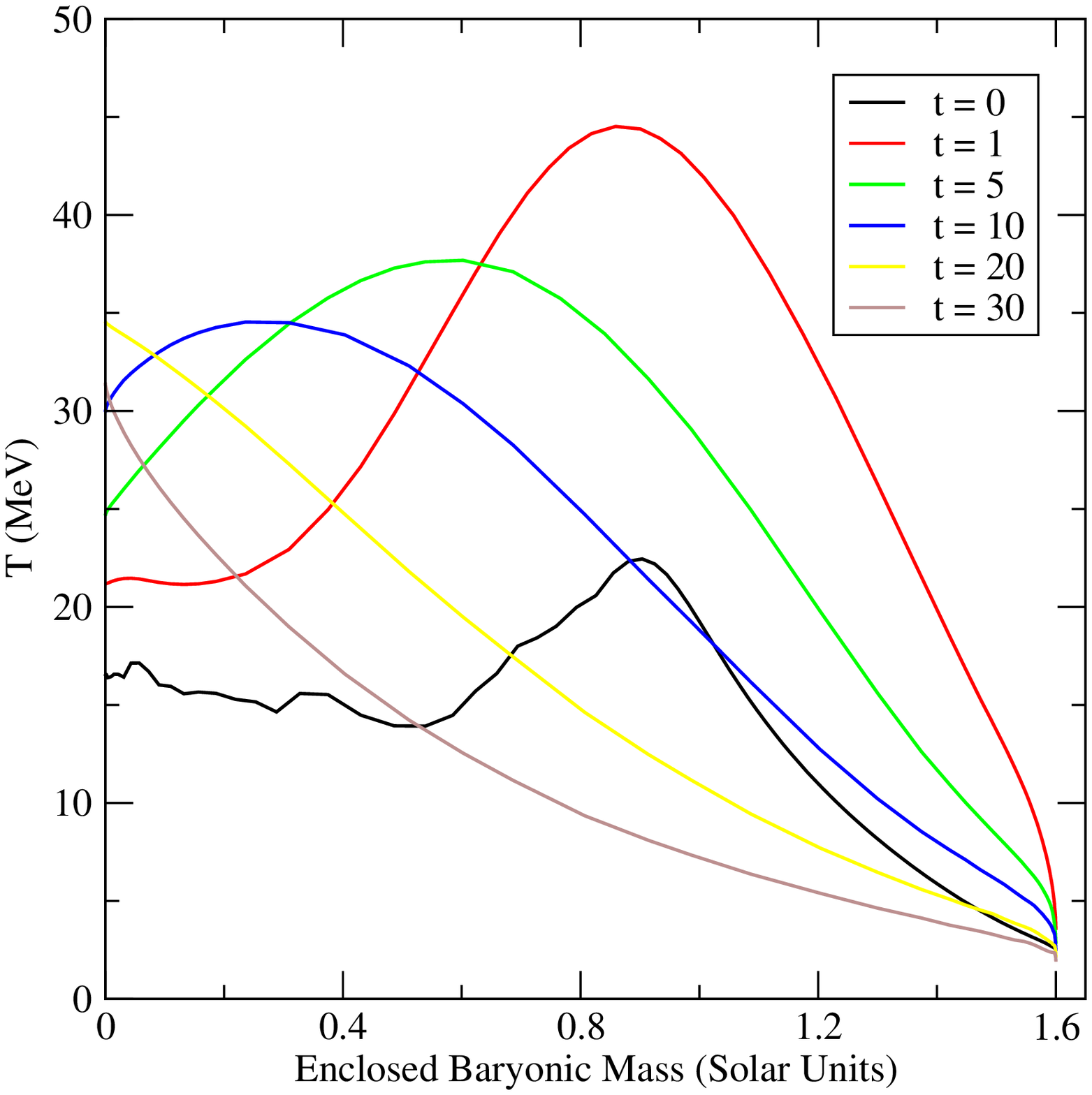}
\includegraphics[width=0.32\textwidth]{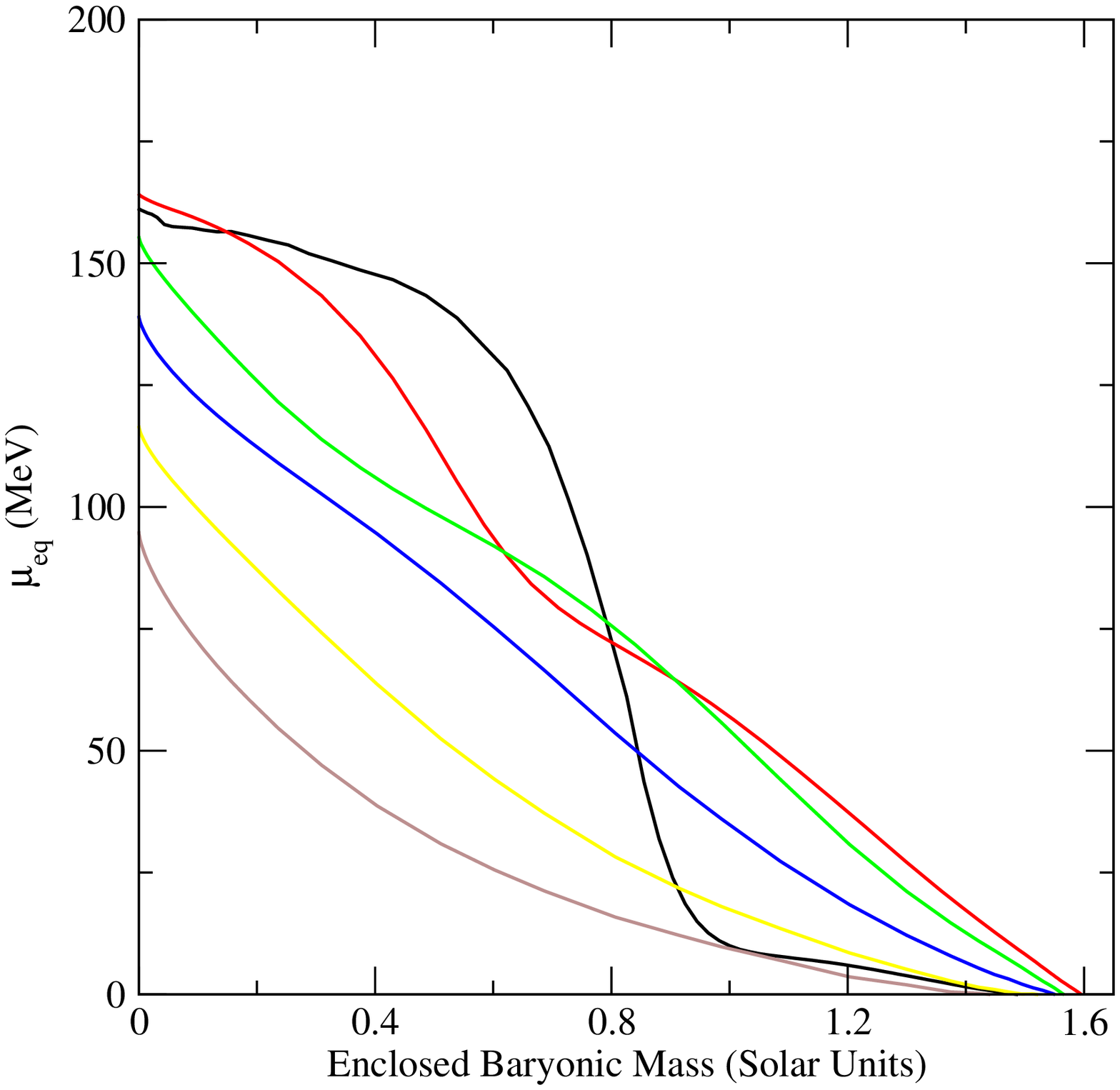}
\includegraphics[width=0.32\textwidth]{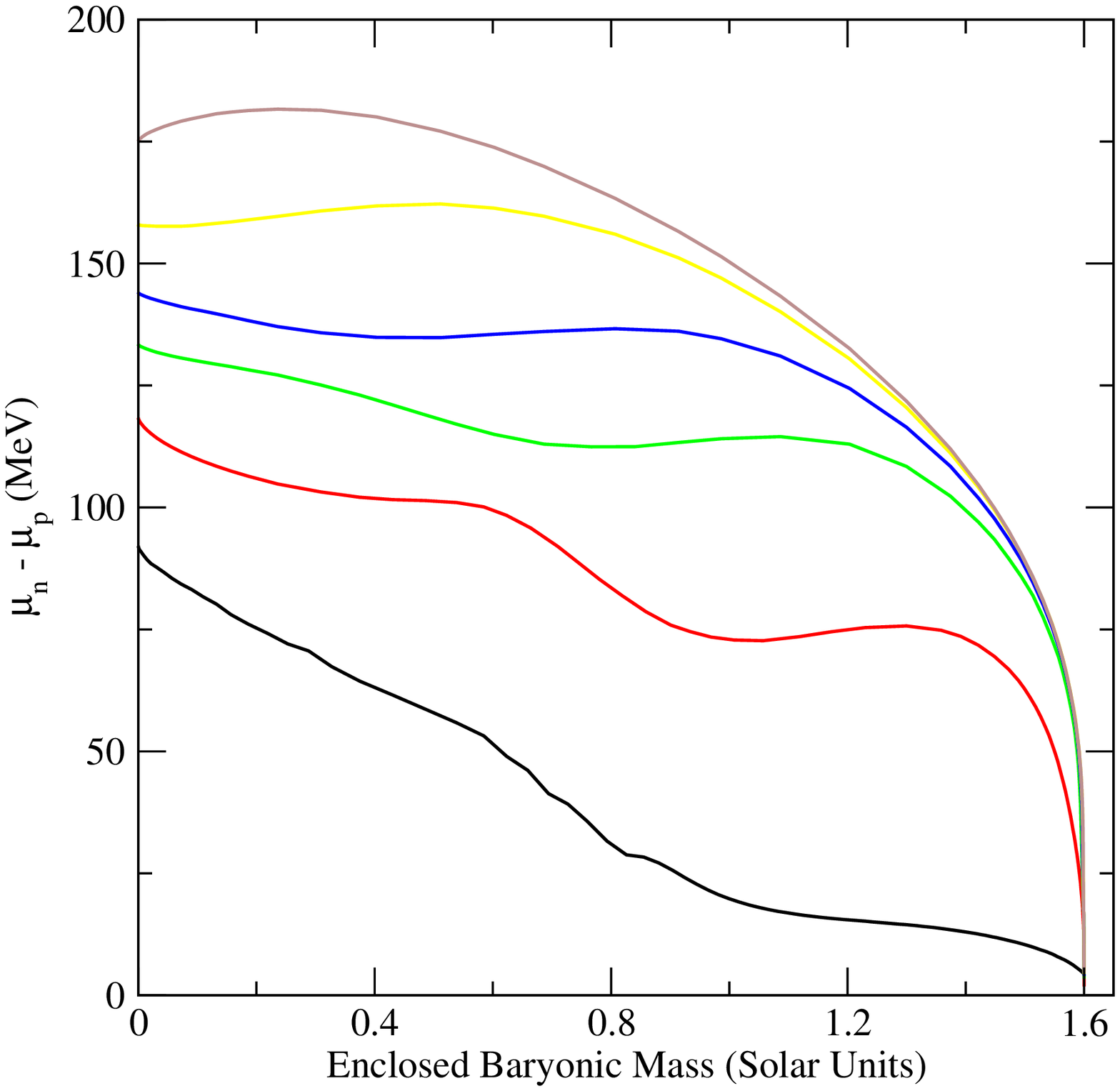}
\includegraphics[width=0.32\textwidth]{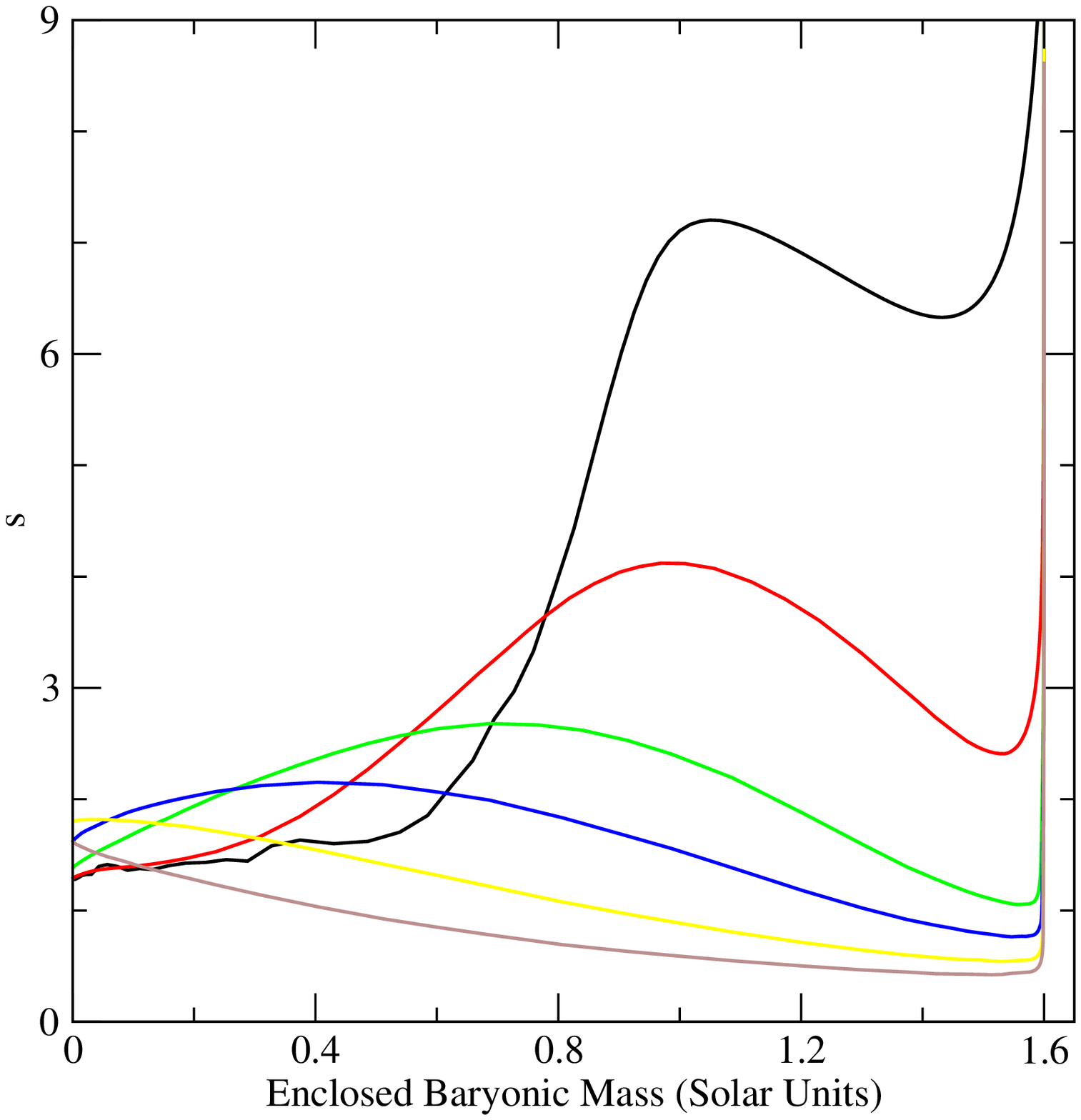}
\includegraphics[width=0.32\textwidth]{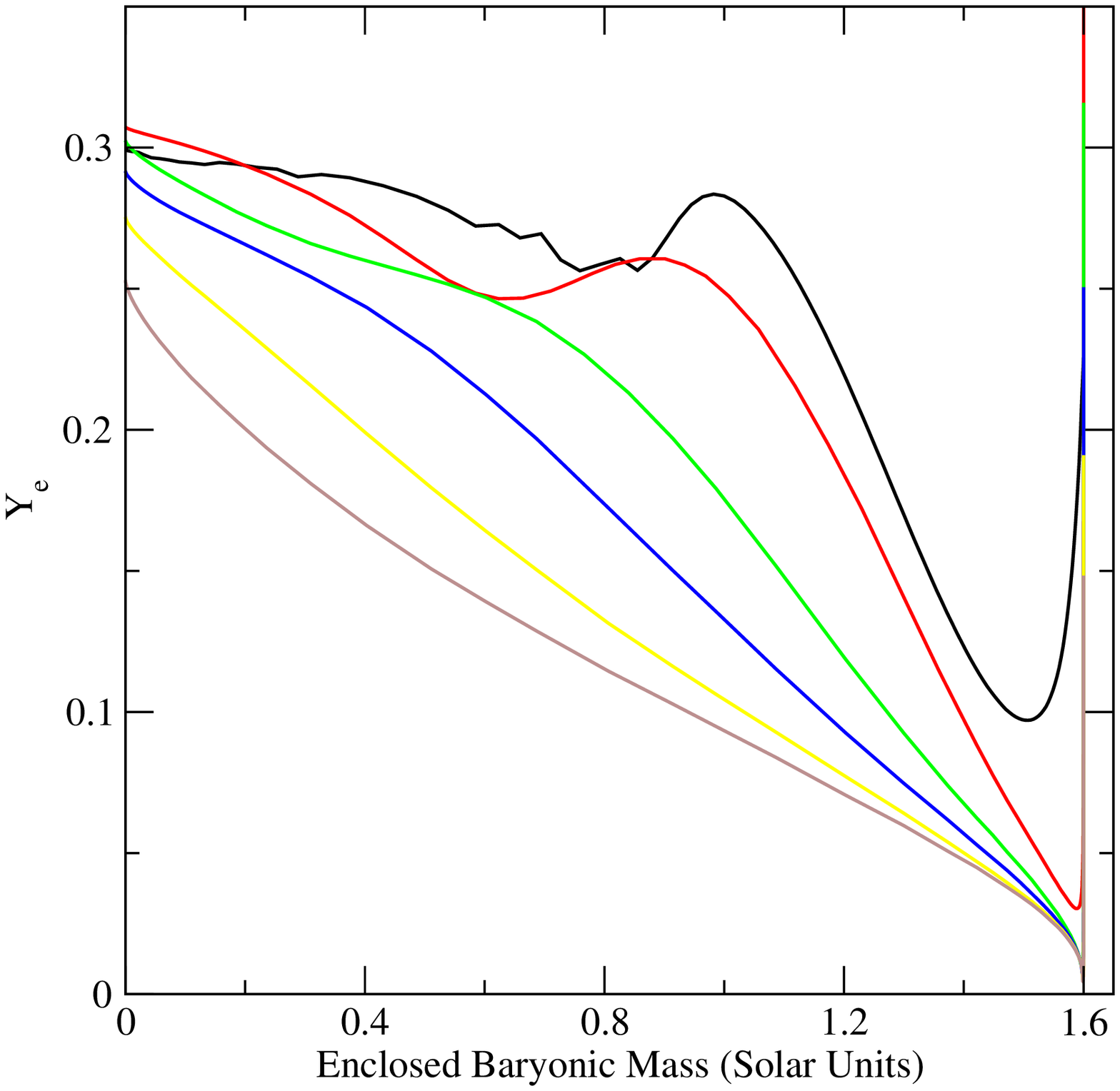}
\includegraphics[width=0.32\textwidth]{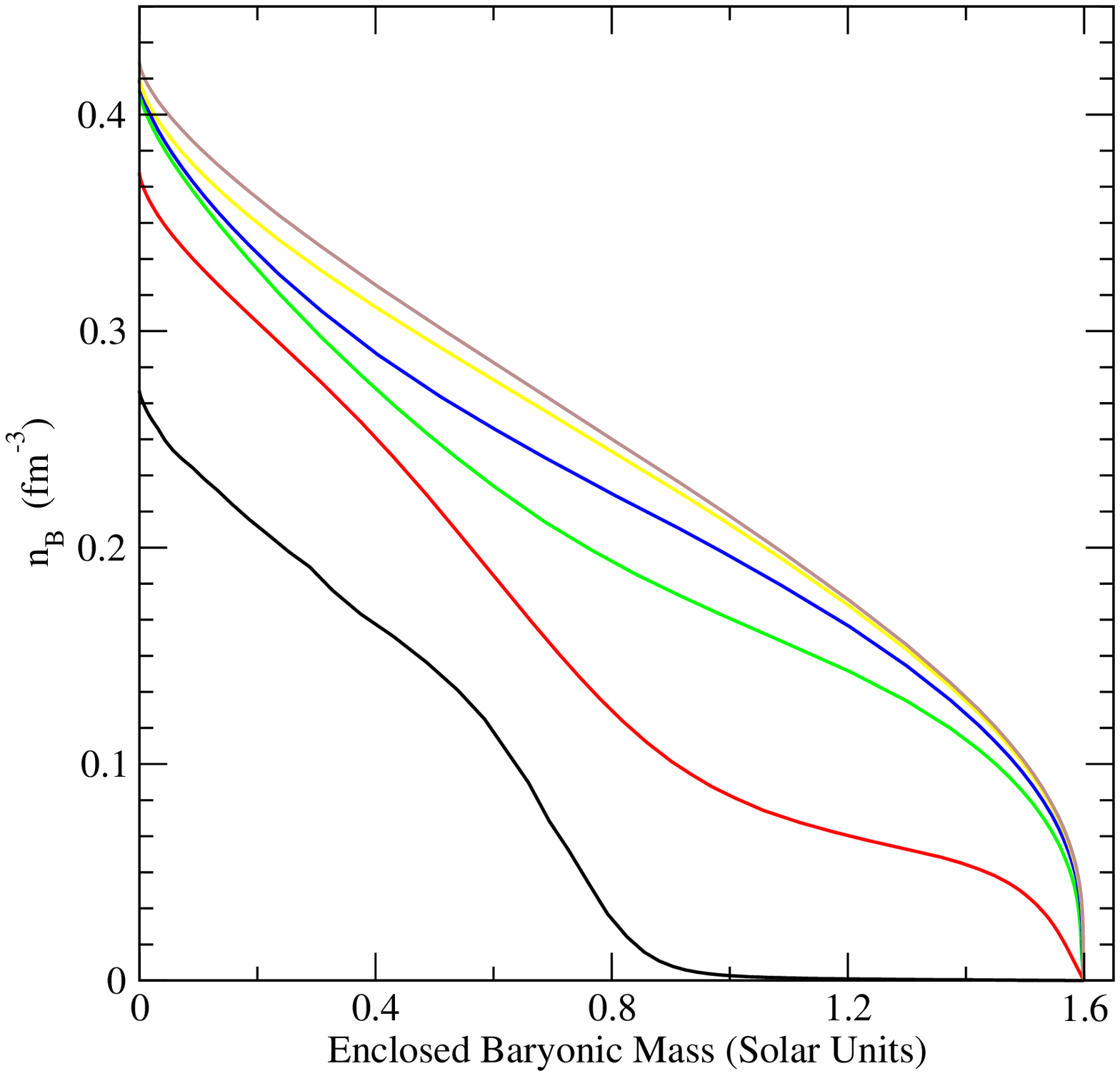}
\caption{The internal structure of the PNS for selected times in the 
fiducial simulation.  The temperature, equilibrium electron neutrino
chemical potential, proton neutron chemical potential difference, 
dimensionless entropy per baryon, electron fraction, and baryon density
are plotted at times 0 s, 1 s, 5 s, 10 s, 20 s, and 30 s, from left to 
right and top to bottom.  The horizontal axes show the enclosed baryon 
number in units of the number of baryons in the sun ($N_\odot \equiv 
\sci{12.04}{56}$).  This figure can be directly compared to figure 
9 of \cite{Pons99}, as it was produced using the same initial model
and a very similar nuclear equation of state and neutrino opacity set.}
\label{fig:Structure}
\end{center}
\end{figure*} 

Rather than follow the collapse of a massive stellar core through
bounce, the calculations here start from a separate calculation of the
highly dynamic phase of initial collapse.  For ease of comparison with
previous work, the $1.6 \, M_\odot$ baryonic mass initial model from
\cite{Pons99} is employed and the affects of convection are not
considered.  This will correspond to a $1.4 \, M_\odot$ gravitational mass
neutron star after it has cooled and can be thought of as 
representative of a standard neutron star \citep{Kiziltan10}.  The
cooling and de-leptonization of this object is followed for 55
seconds, which is shortly after the time the PNS becomes optically
thin.

\subsection{Physical Ingredients}

A relativistic mean field of equation of state consisting of only
neutrons, protons, and electrons is assumed.  The GM3 parameter set is
used without hyperons \citep{Glendenning91}, which is what was used in
\citet{Pons99}.  Neutrino opacities are also calculated in the
relativistic mean field approximation using the formalism of
\citet{Reddy98}.  The tensor polarization is also included so that
``weak magnetism'' affects are included to all orders
\citep{Horowitz03}.  The electron scattering rates from \citet{Yueh77}
are used for the inelastic scattering kernels.  Nucleon scattering is
assumed to occur within a single group, although the opacities are
calculated using the full inelastic differential cross-sections.
Bremsstrahlung is implemented using the structure function given in
\cite{Hannestad98}.  Rather than include this in the annihilation
kernels, the Bremsstrahlung mean free path has been calculated
assuming a thermal distribution for the secondary neutrinos.  Given
the uncertainty in the Bremsstrahlung rate itself and its large
density dependence, this is a reasonable approximation.  Electron
positron pair annihilation \citep{Bruenn85} is also included.  Pure
neutrino processes (i.e.  $\nu_e + \bar \nu_e \rightarrow \nu_\tau +
\bar \nu_\tau$) are not included.  This set of rates is fully
consistent with the rate set used in \cite{Pons99}, but differs
significantly from the rate sets used in recent collapse simulations
\citep{Huedepohl10,Fischer11}.

The study here uses 30 logarithmically spaced energy groups from
2 MeV to 75 MeV plus one final group extending from 75 MeV to 1000 
MeV to encompass the tail of the thermal distribution.  This final 
group is only populated deep in the PNS and it is in tight thermal 
equilibrium due to the extremely short mean free paths for such 
high energy neutrinos.  Minimal differences are found in the PNS
evolution if only 20 groups are employed to cover the same energy
range.

The adaptive radial gridding algorithm is set to keep approximately 
130 zones on the grid and allow for at most a $20\%$ change in 
density across a zone and a $10\%$ change in radius across a zone.  
The boundary pressure is set so that the outer edge of the model has 
a density around $2 \times 10^9 \, {\rm g \, cm}^{-3}$.  This is a 
sufficiently low density that all of the neutrinos have decoupled well
within the outer boundary.

\subsection{Structural Evolution}

Qualitatively, the internal structure of the PNS evolution follows
the standard picture of Kelvin-Helmholtz PNS cooling as described by
\cite{Burrows86}, \cite{Keil95a}, and \cite{Pons99}, where the
gravitational binding energy of the compact object provides energy lost
to neutrino emission.  After the shock produced by the
supra-nuclear density bounce of the core propagates through the outer
layers of the PNS, a high entropy shocked region is left on top of a
cold un-shocked PNS core, which has an entropy similar to the initial
entropy of the pre-supernova iron core.  The outer shocked layers have
de-leptonized during the $\nu_e$ burst, but neutrinos in the core
itself have been trapped since before bounce (although partial
deleptonization has occurred), resulting in a large non-zero
$\mu_{\nu_e,{\rm eq}}$ and $Y_e \approx 0.3$
\citep[c.f.][]{Liebendorfer01b}.  This provides the initial condition
for PNS cooling.

The internal structure of the PNS simulation is shown in figure
\ref{fig:Structure} for a number of times (with time zero
corresponding to the starting point of the simulations, not the time
of core-bounce).  The models start with a core entropy of $\sim 1.2$.
The entropy rises from 1.6 at an enclosed baryonic mass of $\sim 0.6
\, M_\odot$ to 7.4 at an enclosed mass of $1.0 \, M_\odot$.  This
implies that the supernova shock was born at around $0.6 \, M_\odot$,
which is reasonably consistent with the core-collapse results of
\cite{Thompson03}. The shocked mantle is at low density relative to
the core and extends to large radius (material that is at a density of
$10^{-5} \, {\rm fm}^{-3}$ is found at $99 \, {\rm km}$), mainly due
to the thermal contribution to the pressure.

From this initial state, the shock heated mantle rapidly contracts
over the first second or so of the simulation.  This contraction is
driven by the rapid loss of energy and lepton number via neutrinos,
which can readily escape due to the low density of the envelope and
long interaction mean free paths.  The loss of lepton number and
thermal energy reduces pressure support in the mantle, and the mantle
responds by rapidly contracting (i.e., rapid relative to the cooling
timescale of the core, not rapid compared to the dynamical timescale
of the envelope).  By two seconds into the simulation, material at a
density of $10^{-5} \, {\rm fm}^{-3}$ is at $17 \, {\rm km}$.  This is
fairly close to the cold neutron star radius for GM3 ($13.5 \, {\rm
  km}$).  The work provided by this contraction is enough to increase
the peak temperature of the mantle from $22 \, {\rm MeV}$ to $45 \,
{\rm MeV}$ even though the entropy of the mantle has decreased from 7
to 4 over this period.

This period of the PNS evolution is most likely
to be sensitive to the initial conditions for the simulations, as 
at later times the details of the initial structure should 
be washed out.  The envelope of the PNS should also be
convective, which significantly alters the rate of energy and lepton 
number transport in the PNS \citep[c.f.][]{Roberts12}.  Additionally,
there might be significant accretion luminosity over this period 
(although this is approximately accounted for by the mantle).  
Therefore, especially given the older provenance of the initial 
conditions, the results from this period should be taken as only 
qualitatively correct.

While the mantle is contracting, $\bar \nu_e$s and $\nu_x$s are 
being transported down the positive radial temperature gradient 
into the core while the $\nu_e$s are being transported outwards
down the large equilibrium chemical potential gradient.  This 
results in a net heat flux into the core and a net lepton flux
out of the core.  This has been referred to as ``Joule heating''  
of the core in previous work \citep{Burrows86}.  Additionally, the 
inner regions contract over this period due to the increased 
boundary pressure on the un-shocked core from the cooling mantle.
This contributes to the temperature increase in the core in addition
to the Joule heating.

After the initial period of mantle contraction, the density structure
of the PNS becomes similar to that of a cold PNS.  Joule heating
continues to increase the temperature of the inner most regions until
the central temperature reaches its peak value of $35 \, {\rm MeV}$ at
$18 \, {\rm s}$ in the simulation.  Then, the temperature of the
entire star falls with time.  The entropy evolution exhibits a similar
behavior.  Lepton number is lost from the entire PNS core over this
time and the electron fraction evolves toward the expected value for
matter in beta-equilibrium with no net electron neutrino number.
After about 15 seconds, contraction slows since the PNS is nearly at
the cold neutron star radius. After this, neutrino emission is powered
chiefly by the loss of thermal energy from the star.

It is also worth noting that the temperature gradient and the
$\mu_{\nu_e,{\rm eq}}$ gradient in the shocked layers of the PNS
become increasingly shallow from 1 s onwards.  Additionally, as the
density of the outer layers rises, the neutron proton chemical
potential difference $\hat \mu = \mu_n - \mu_p$ gets larger.  The
increase in $\hat \mu$ and the decrease in $\mu_{\nu_e,{\rm eq}}$
bring $\hat \mu$ close to the electron chemical potential $\mu_e =
\hat \mu + \mu_{\nu_e,{\rm eq}}$ as time goes on.  These
considerations have significant consequences for the spectral
evolution of the neutrinos and which are discussed in section
\ref{sec:Spectra Compare}.

\subsection{Emergent Luminosity and Spectral Evolution}

\begin{figure}
\leavevmode
\begin{center}
\includegraphics[width=\columnwidth]{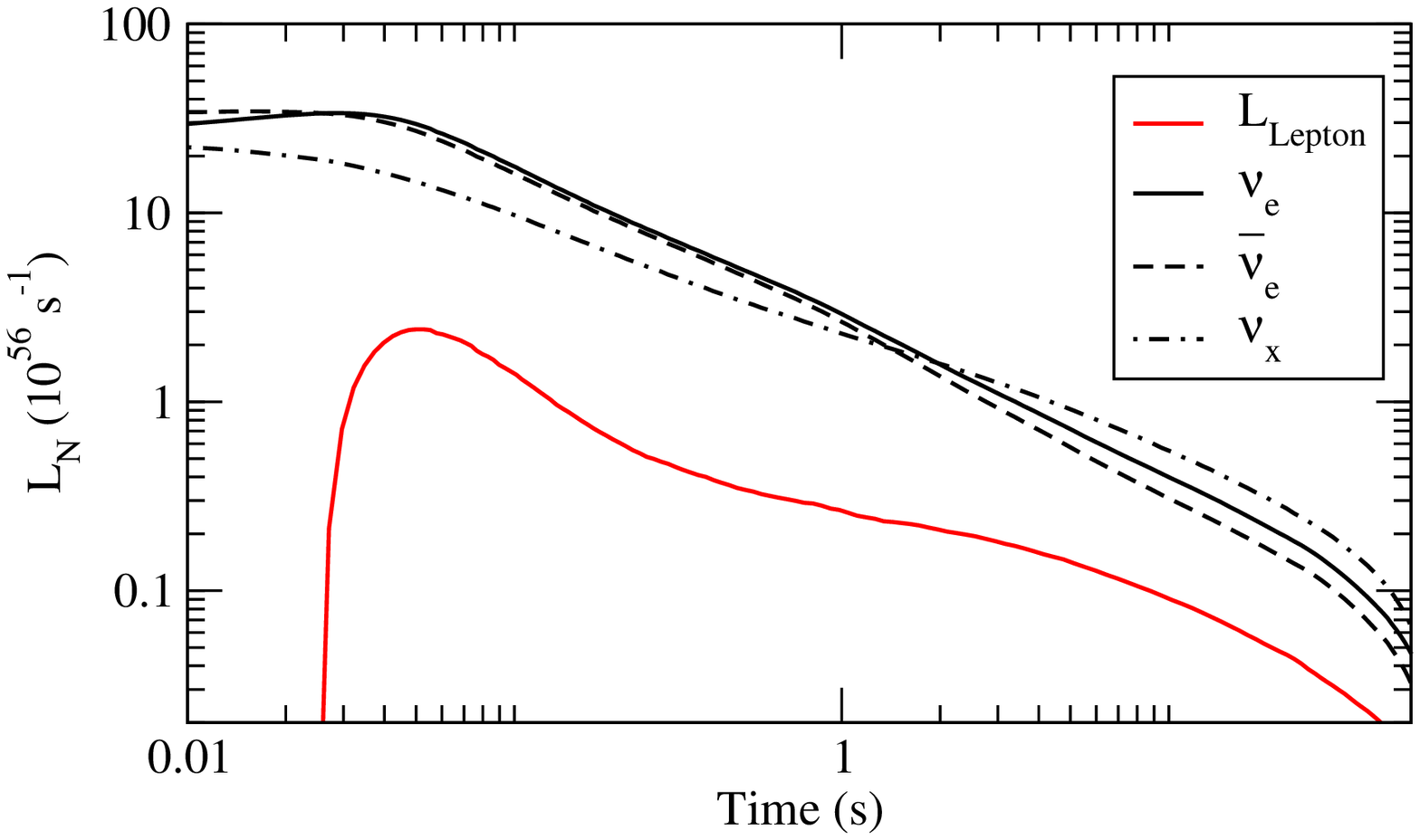}
\includegraphics[width=\columnwidth]{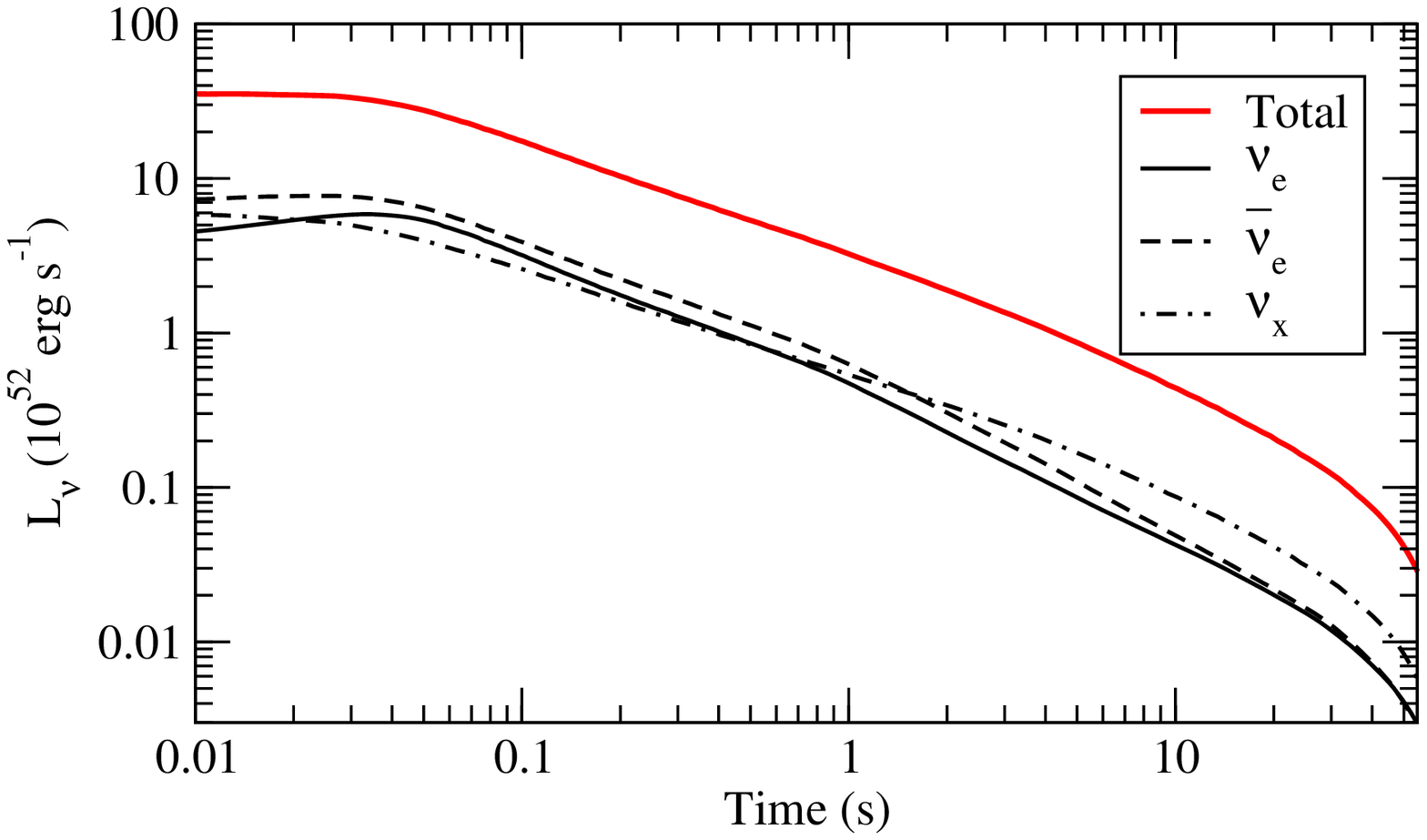}

\caption{Top panel: Number luminosities as a function of time for 
$\nu_e$ (solid black line), $\bar \nu_e$ (dashed black line), 
$\nu_x$ (dot-dashed black line) and the de-leptonization
rate, $\dot N_{\nu_e} - \dot N_{\bar \nu_e}$. Bottom panel: Energy 
luminosities as a function of time.  The black lines are the same
as in the top panel, but the solid red line is the total energy 
emitted in neutrinos per time.}
\label{fig:Luminosities}
\end{center}
\end{figure} 

The total integrated energy loss in neutrinos over the duration of 
the simulation is $E_\nu = \sci{2.32}{53} \, {\rm erg}$, and the 
total lepton number radiated is $N_L = \sci{3.2}{56}$.
The neutrino emission from the PNS is shown in figure 
\ref{fig:Luminosities}.  As is discussed above, the first couple 
of seconds are dominated by the contraction of the PNS mantle. 
Over the first second of the simulation, $38\%$ of the total 
neutrino energy loss and $20\%$ of the total lepton number loss occurs. 
During this period, the $\nu_x$ number luminosity is produced 
mainly by the un-shocked core, as the $\mu$ and $\tau$ neutrinos 
are mainly coupled to the envelope through scattering.  Therefore, 
the luminosity in these flavors is lower because of the smaller
emitting surface (which is not offset by the temperature of 
the core).  

During mantle contraction, there is a high de-leptonization rate 
driven by the outermost layers of the star.  After 
the first few hundred milliseconds, de-leptonization
slows as the outer layers go towards $\mu_{\nu_e,{\rm eq}} \approx 0$ 
and de-leptonization is driven by diffusion out of the core.  The 
values of the de-leptonization rate before 80 ms are unrealistic, as 
they are determined by the relaxation of the assumed initial conditions 
for the neutrinos. 

\begin{figure}
\leavevmode
\begin{center}
\includegraphics[width=\columnwidth]{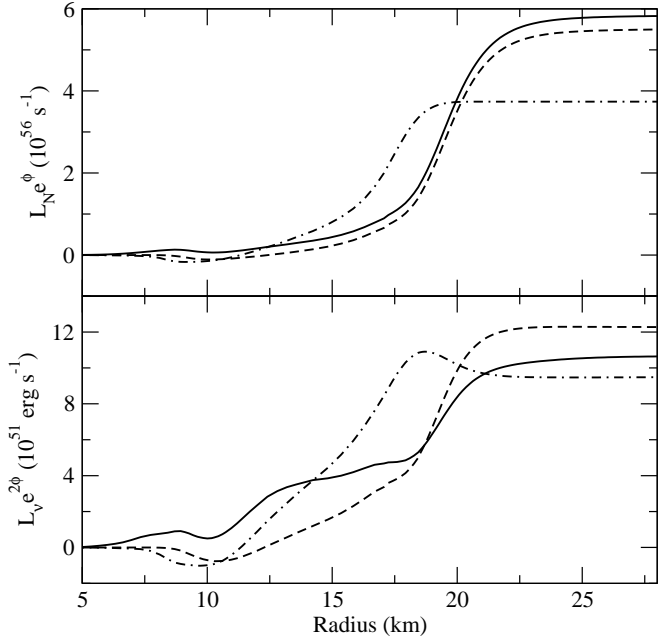}
\caption{Neutrino number and energy luminosities at infinity as a function 
of radius at 500 ms into the simulation.  The solid lines are for electron 
neutrinos, the dashed lines are for electron antineutrinos, and the 
dot-dashed lines are for $\mu$ and $\tau$ neutrinos.}
\label{fig:radial_flux}
\end{center}
\end{figure} 

Over the first two seconds, the $\nu_x$ luminosities are significantly
lower than the luminosities of the electron flavored neutrinos.  
The neutrino energy and number luminosities as a function of 
radius at 500 ms after the beginning of the simulation are 
shown in figure \ref{fig:radial_flux}. First, this illustrates 
that the $\mu$ and $\tau$ neutrino number fluxes are 
being set much further inside the star (at around 18 km) than the electron 
neutrinos, but they exchange energy out to a significantly larger radius via 
scattering.  Second, there is an inward directed anti-electron, $\mu$ and 
$\tau$ flux near the mantle core boundary.  As cooling precedes, heat diffuses 
down the positive temperature gradient (and positive equilibrium chemical 
potential gradient for the anti-electron neutrinos) into the lower entropy core.
This is the Joule heating discussed above.  In contrast, the large negative 
equilibrium chemical potential gradient for the electron neutrinos overwhelms 
the positive radial temperature gradient and the electron neutrino flux is 
positive everywhere. 

After the PNS has contracted to close to the cold neutron star radius, 
the $\nu_x$ luminosity has increased relative to the $\nu_e$ and $\bar 
\nu_e$ luminosities.  In fact, the $\nu_x$ luminosity is about twice 
the luminosity in either of the electron neutrino species.  These neutrinos
decouple further inside the PNS and are therefore emitted at a higher 
effective temperature, resulting in a larger number and energy luminosity.
Between thirty and forty seconds the PNS becomes transparent to neutrinos
and the luminosity drops off significantly.

\begin{figure}
\begin{center}
\leavevmode
\includegraphics[width=\columnwidth]{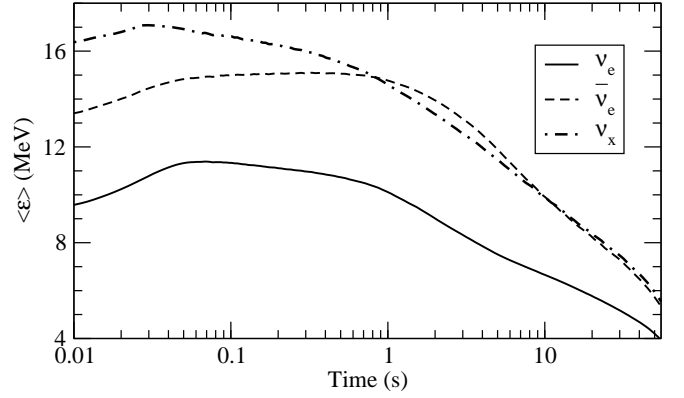}
\caption{Average energies of the emitted neutrinos measured at infinity.}
\label{fig:Average Energies}
\end{center}
\end{figure}

The average energies of the emitted neutrinos at infinity as a function of 
time are shown in figure \ref{fig:Average Energies}.  Within the integrated 
energy group formalism, the neutrino energy moments at infinity are defined 
as
\be
\langle \epsilon^n \rangle = e^{n\phi_s} \frac{\sum_g \langle 
\omega \rangle_g^{n-1} H_g}{\sum_g F_g},
\ee
where $\langle \omega \rangle_g$ is a group averaged energy and $\phi_s$ is the 
surface value of the metric potential.  

During the mantle contraction phase, there is the standard hierarchy
of neutrino average energies $\langle \epsilon_{\nu_e} \rangle <
\langle \epsilon_{\bar \nu_e} \rangle < \langle \epsilon_{\nu_x}
\rangle$.  After mantle contraction has ceased, the energy decoupling
radius of electron neutrinos and $\mu$ and $\tau$ neutrinos becomes
similar and for the rest of the PNS evolution $\langle \epsilon_{\bar
  \nu_e} \rangle \approx \langle \epsilon_{\nu_x} \rangle$.  This is
in contrast to the difference between the electron neutrino and
anti-neutrino average energies, which obey $\langle \epsilon_{\nu_e}
\rangle < \langle \epsilon_{\bar \nu_e} \rangle$ for the entire
calculation, although the two average energies get closer at late
times.  An analysis of why this is, its implications, and a comparison
to other results in the literature is given in section
\ref{sec:Spectra Compare}.

\begin{figure}
\begin{center}
\leavevmode
\includegraphics[width=\columnwidth]{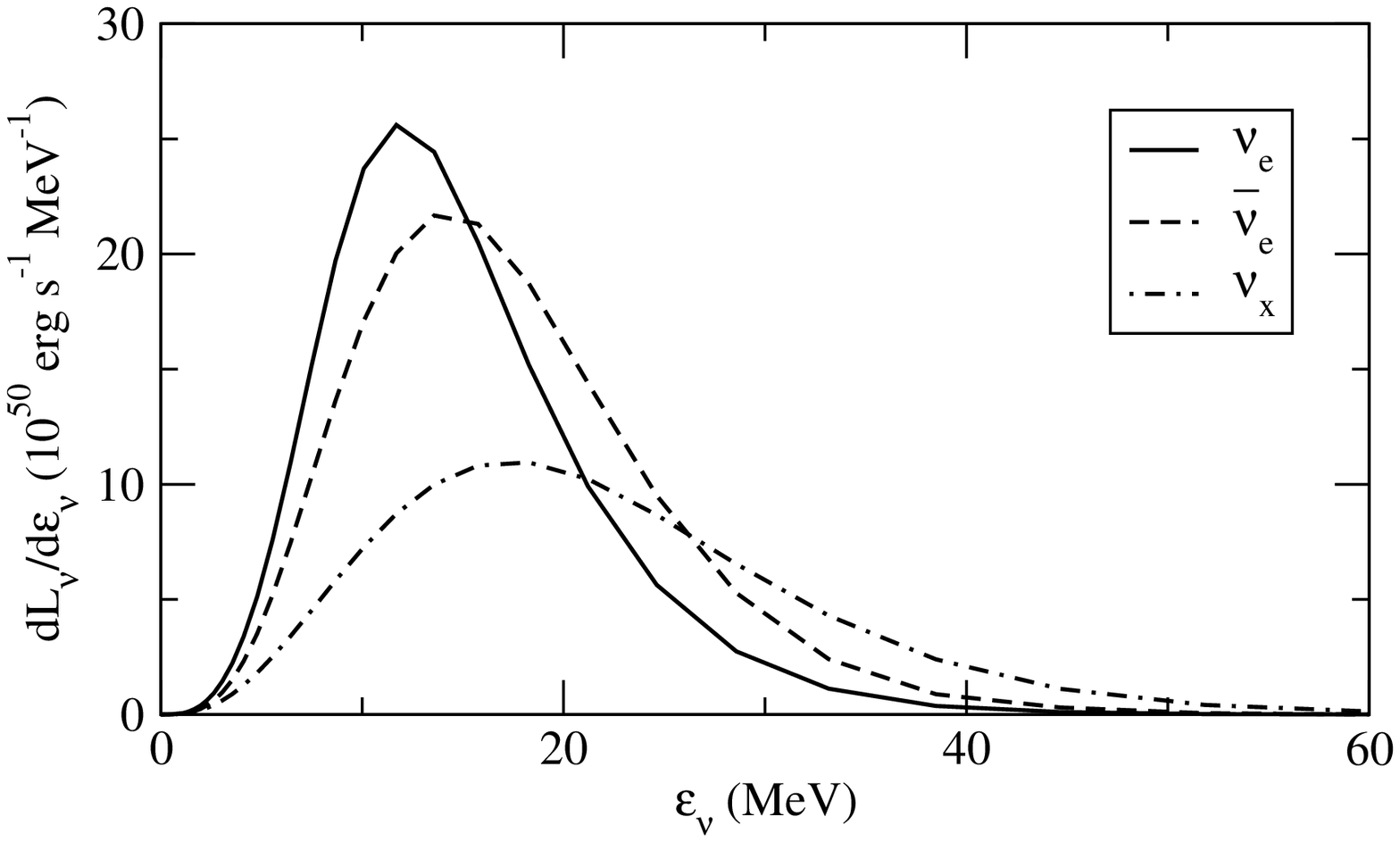}
\includegraphics[width=\columnwidth]{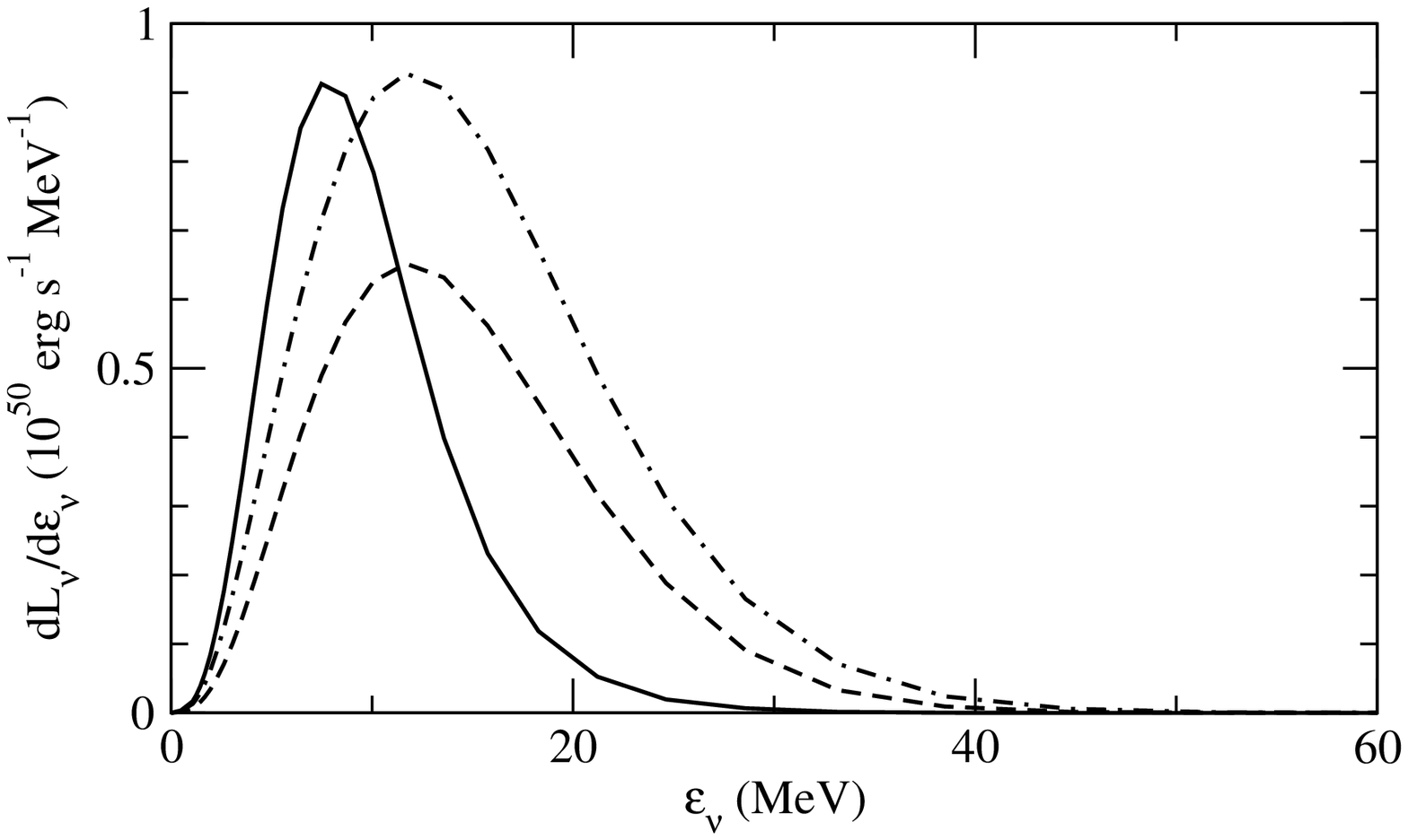}
\caption{Neutrino spectra for all three flavors tracked 
in the simulation.  The top panel is at 100 ms after the 
start of the simulation, the bottom panel is the spectrum
at 5 seconds.}
\label{fig:Spectra}
\end{center}
\end{figure}
The emitted neutrino spectra at two representative times are 
shown in \ref{fig:Spectra} for reference.  The $\nu_x$ neutrinos 
decouple further in the star than the $\bar \nu_e$ neutrinos at 
five seconds into the simulation so that they have a larger number 
luminosity due to the larger temperatures found there, but these 
flavors have a similar energy sphere due to inelastic scattering 
which accounts for the similar value of the peak of the luminosity
as a function of neutrino energy.

\section{Discussion}
\label{sec:Discussion}

\subsection{Comparison to EFLD}
\label{sec:EFLDComp}

Until recently \citep{Huedepohl10,Fischer10}, most studies of PNS
cooling used the EFLD approximation to describe neutrino transport
\citep{Burrows86,Keil95a,Pons99,Roberts12}.  It is thus a worthwhile
exercise to compare the results obtained using EFLD and the present
variable Eddington factor method for transport.  No detailed
comparison of the effect of different flux limiters is attempted since
EFLD clearly breaks down in the decoupling regime independent of the
flux limiter used.  See \citet{Messer98} and \citet{Pons00b} for
discussions of the affect of different flux limiters.

\begin{figure}
\leavevmode
\begin{center}
\includegraphics[width=\columnwidth]{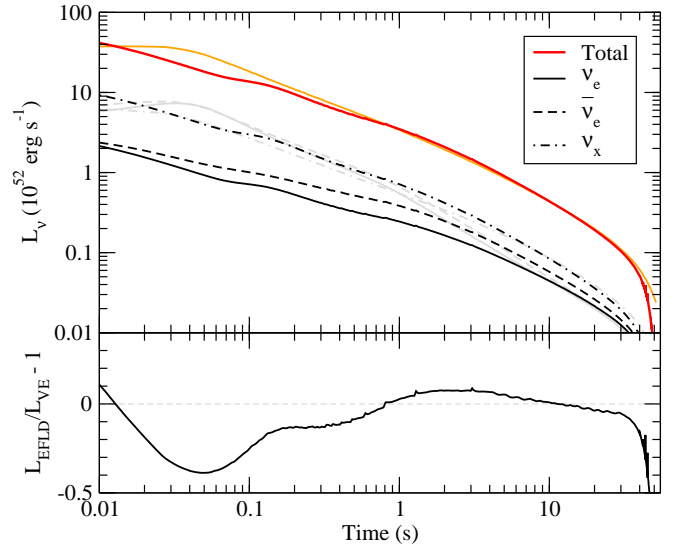}
\caption{Luminosities as a function of time for $\nu_e$ (solid black 
line), $\bar \nu_e$ (dashed black line), $\nu_x$ (dot-dashed black line) 
of time and the total luminosity (red line) in the EFLD approximation.
The gray and orange lines is the data from figure \ref{fig:Luminosities}. 
The bottom panel shows the ratio of the total EFLD luminosity to the 
total luminosity calculated using the new code.}
\label{fig:EFLD_Luminosities}
\end{center}
\end{figure}

Models using the identical microphysics and initial conditions
described in section \ref{sec:PNSEvolution} were calculated using the
EFLD code described in \citet{Roberts12} with convection turned
off. This code is similar to the one used in \cite{Pons99} and is
completely different from the one used in this paper.  A derivation of
EFLD in the context of this paper is included in Appendix
\ref{sec:EFLD}.  The EFLD luminosities as a function of time
are shown in figure \ref{fig:EFLD_Luminosities}, alongside the
luminosities from section \ref{sec:PNSEvolution}.  EFLD clearly does a
reasonably good job of predicting the total neutrino luminosity, but
poorly predicts the luminosities of each flavor.

At early times, some deviation in the total luminosity is expected because 
the mantle, which is driving most of the neutrino emission,
is not particularly optically thick.  Additionally, at late times when the 
whole PNS becomes optically thin, EFLD deviates from the variable Eddington
factor solution.  But for the bulk of the PNS evolution the deviation 
between the two methods is around $10\%$, which is surprisingly good 
agreement.  Given that most predictions made using EFLD codes have relied
only on the total neutrino luminosity, it seems that previous results can
be reasonably trusted.  The total luminosity emitted from the PNS is set 
at the neutrino spheres of each flavor, which is the last point at which
EFLD can be considered reliable.  The outermost layers of the PNS in
which the neutrinos decouple can come into radiative equilibrium on a short
timescale and therefore rapidly adjust to the flux being pushed through them 
from below.  As neutrinos propagate through the outer layers in the EFLD 
formalism, flux may be shifted between flavors unrealistically but the outer
layers of the PNS will rapidly evolve to carry the right total luminosity.  
Therefore, it is not surprising that EFLD gets the total luminosity right 
but fails to predict the luminosities of specific flavors.  Of course, EFLD 
makes no predictions regarding the spectral properties of the neutrinos.  

\subsection{Neutrino Spectra and The Composition of The Neutrino Driven Wind}
\label{sec:Spectra Compare}

The most striking difference between the present simulations and other
recent studies \citep{Huedepohl10,Fischer10} is the greater difference
in the present study of the electron neutrino and anti-neutrino
average energies at late times.  There are a number of possible reasons
for this difference.

One is the initial model chosen for the PNS evolution.  Rather than
use an initial model from a separate calculation of core-collapse,
both \cite{Huedepohl10} and \cite{Fischer10} follow the entire
evolution of the supernova.  In so far as the initial models are
similar, the two approaches should give the same answer.  The initial
model used here is somewhat dated and was chosen mainly to facilitate
the comparison with the work of \cite{Pons99}.  At early times the
initial progenitor model will certainly affect the properties of the
emitted neutrinos significantly, but after the first second
\citet{Pons99} found that the evolution does not depend sensitively on
the initial progenitor model.  Of course, the difference in the
average energies of the electron and anti-electron neutrinos is a
fairly subtle effect.  Therefore, the effect of the progenitor model
should not be ruled out, but based on the argument below it seems 
unlikely that the progenitor model is the dominant factor.

It is possible that the methods used for transport differ 
enough to give disparate results.  This also seems 
unlikely considering all three approaches come close to directly 
solving the Boltzmann equation, that the formalism described in
this work is fairly similar to the formalism of \cite{Huedepohl10} 
\citep[see][]{Rampp02}, and that the approaches of \cite{Fischer10}
and \cite{Huedepohl10} have been shown to yield similar results 
\citep{Liebendorfer05}.  

A more significant difference though may be the microphysics
employed.  The difference between the electron neutrino and
anti-neutrino spectral temperatures is mainly set by the difference
between their respective mean free paths to capture on nucleons, as
the scattering mean free paths for both species are nearly equal.  Due
to de-leptonization, there are far more neutrons to capture electron 
neutrinos than protons to capture electron anti-neutrinos.  Of course, 
it is possible for both of these reactions to have strong final state 
blocking (electron blocking for the neutrinos and neutron blocking for the
anti-neutrinos).  If it is assumed that there is no energy transfer to
the nucleons, as in \cite{Fischer11}, then both reactions will be
strongly blocked, the elastic interaction rates described in
\cite{Bruenn85} go to the same value, and it is expected that average
electron neutrino and anti-neutrino energies will be similar at late 
times due to the similar charged current mean free paths for both species.

\begin{figure}
\leavevmode
\begin{center}
\includegraphics[width=\columnwidth]{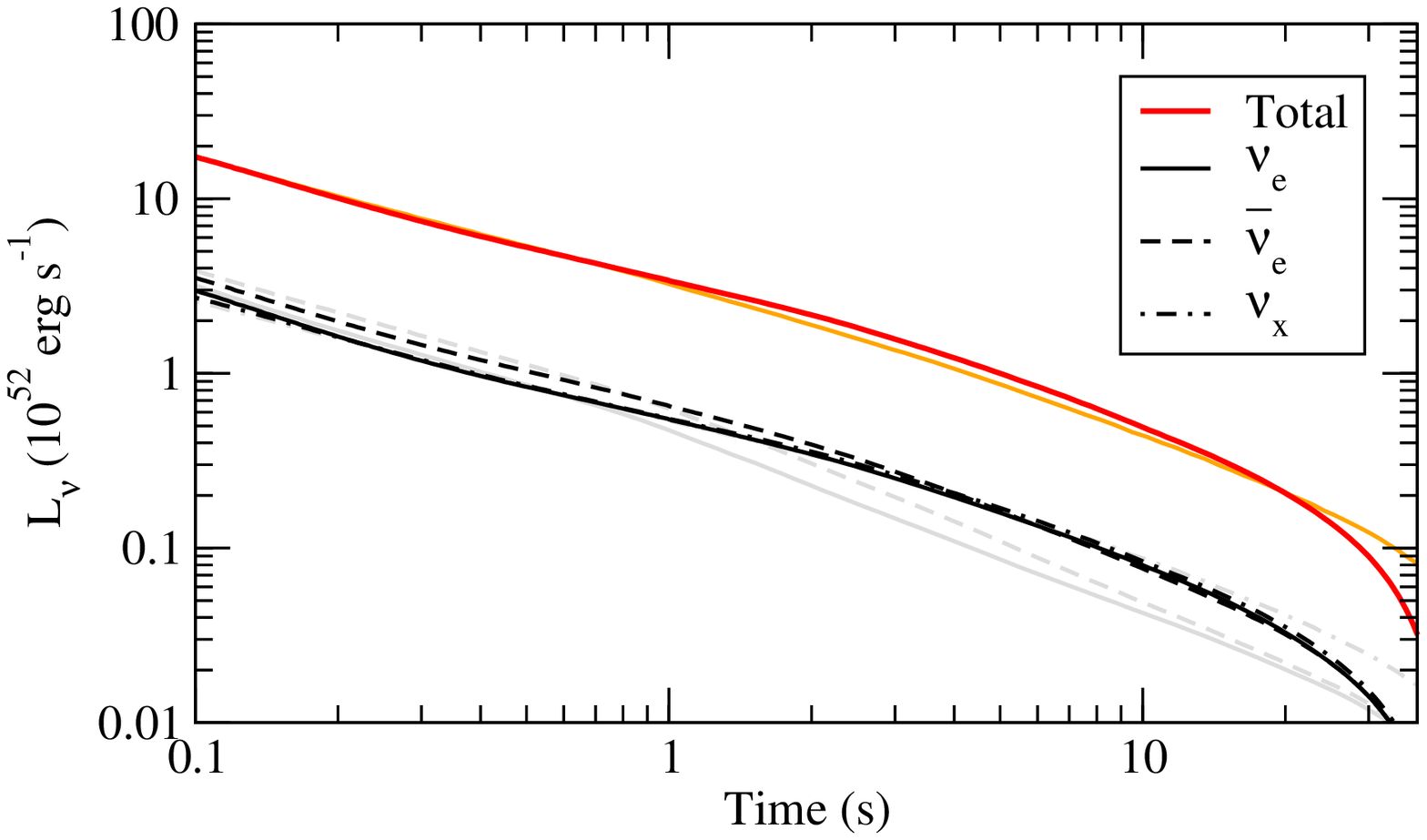}
\includegraphics[width=\columnwidth]{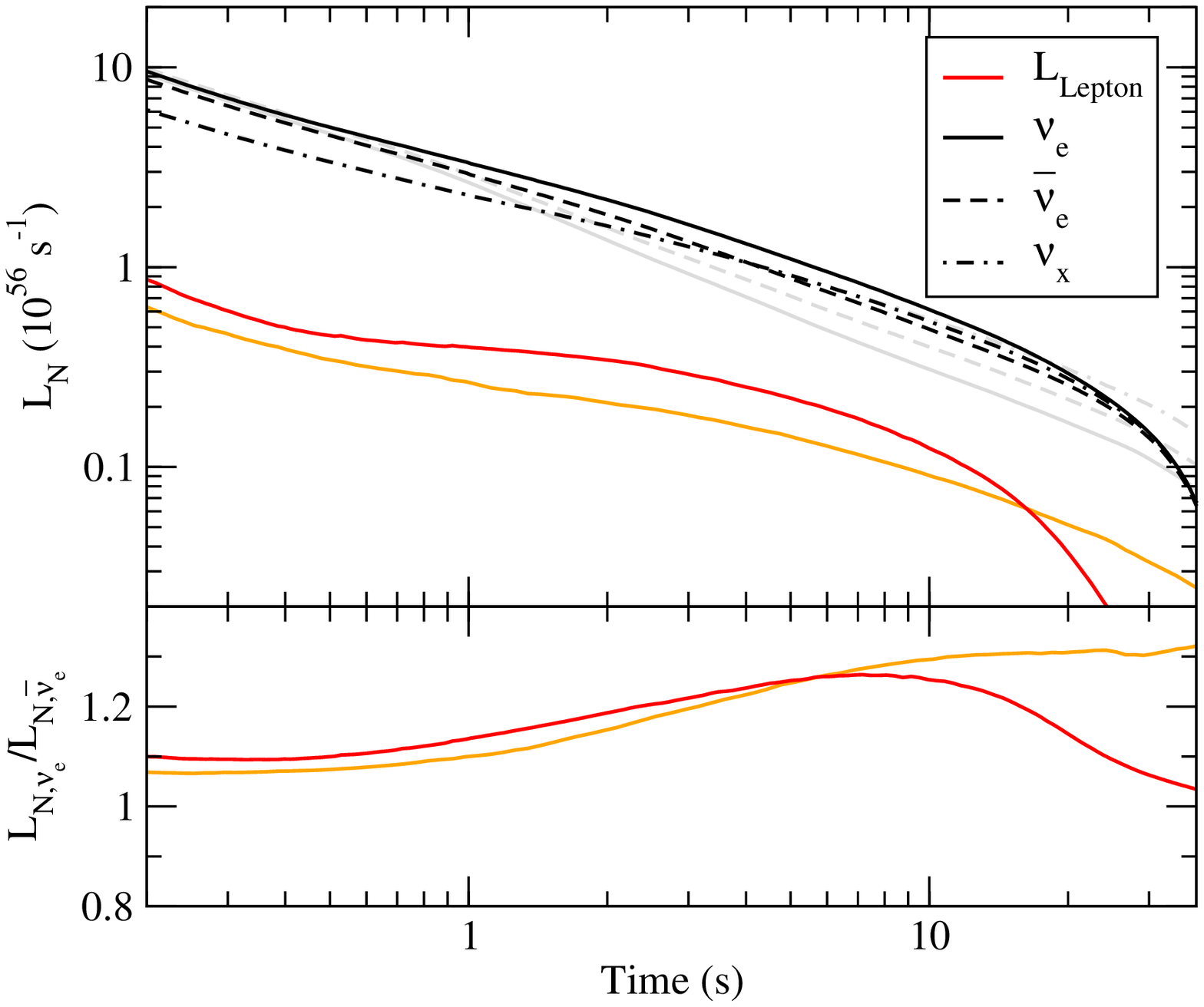}
\caption{Energy and number luminosities for a model using the 
\cite{Bruenn85} nucleon capture rates compared to the fiducial model 
described in section \ref{sec:PNSEvolution}.  Notice the slightly 
increased $\nu_e$ and $\bar\nu_e$ cooling rates at late times and 
the convergence of all three luminosities.  The black lines 
and red lines are for the model using \cite{Bruenn85} rates, while
the gray and orange lines are for the model described in section
\ref{sec:PNSEvolution}.  In the bottom plot, the ratio $\dot 
N_{\nu_e} /\dot N_{\bar \nu_e}$, which is of consequence to the 
electron fraction in the neutrino driven wind, is shown in the 
bottom plot of the second panel.}
\label{fig:Luminosity_compare}
\end{center}
\end{figure} 

The final state blocking symmetry predicted by the charged current rates 
of \citet{Bruenn85} does not agree with more detailed calculations 
of the electron neutrino capture rates.  There is in fact a strong asymmetry 
between the two reactions, because there is significantly more energy 
available in the entrance channel for $\nu_e + n \rightarrow e^- + p$ than
for $\bar \nu_e + p \rightarrow e^+ + n$.  The difference between 
the energy of the entrance channels is just the difference between the 
fermi energies of the neutrons and protons.  The Fermi energies for interacting
nucleons are given by $e_{F,i} = k^2_{F,i}/2M_i + U_i$, where $U_i$ is an 
isospin dependent potential energy due to strong interactions in the medium.
For neutron rich conditions, the neutron potential energy is larger than the 
proton potential energy due to the nuclear symmetry energy.  Most of the potential
difference, $U_N-U_P$ is transferred to the outgoing electron in the reaction 
$\nu_e + n \rightarrow e^- + p$.  This effect can significantly decrease the 
absorption mean free path for electron neutrinos. Due to the large value of the nuclear 
symmetry energy relative to the value expected for free nucleons, $U_N-U_P$ 
accounts for a significant fraction of $\hat \mu$.  Although this amount of 
energy is often not enough to put the final state electron above the electron 
Fermi surface, it is enough to put the final state electron in a relatively less 
blocked portion of phase space.  This effect is included in the relativistic 
formalism of \cite{Reddy98}, which is used to calculate the neutrino interaction
rates used in the models presented in this work.  The details of the importance
of realistic kinematics on charged current rates will be discussed in future work.

\begin{figure}
\begin{center}
\leavevmode
\includegraphics[width=\columnwidth]{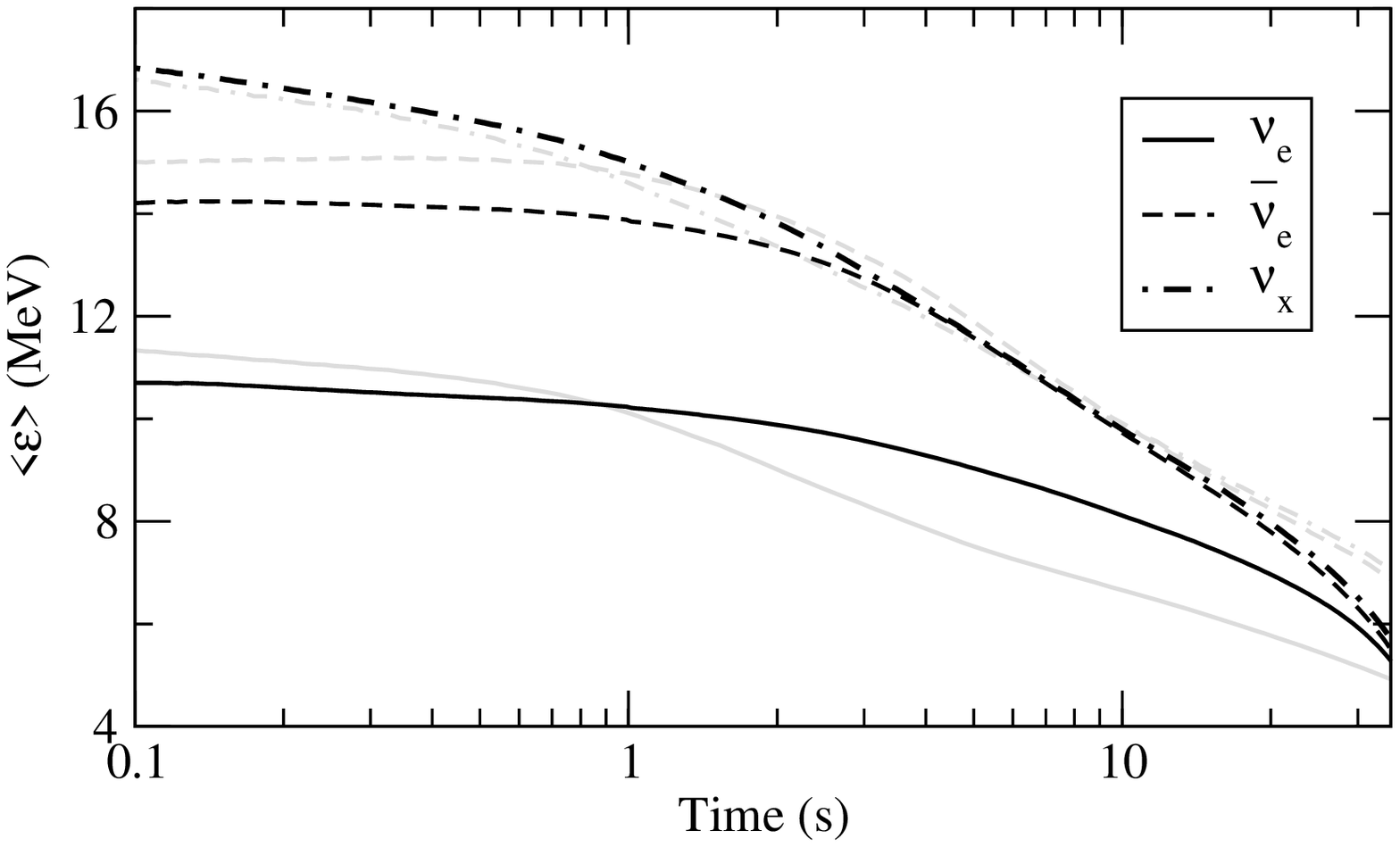}
\includegraphics[width=\columnwidth]{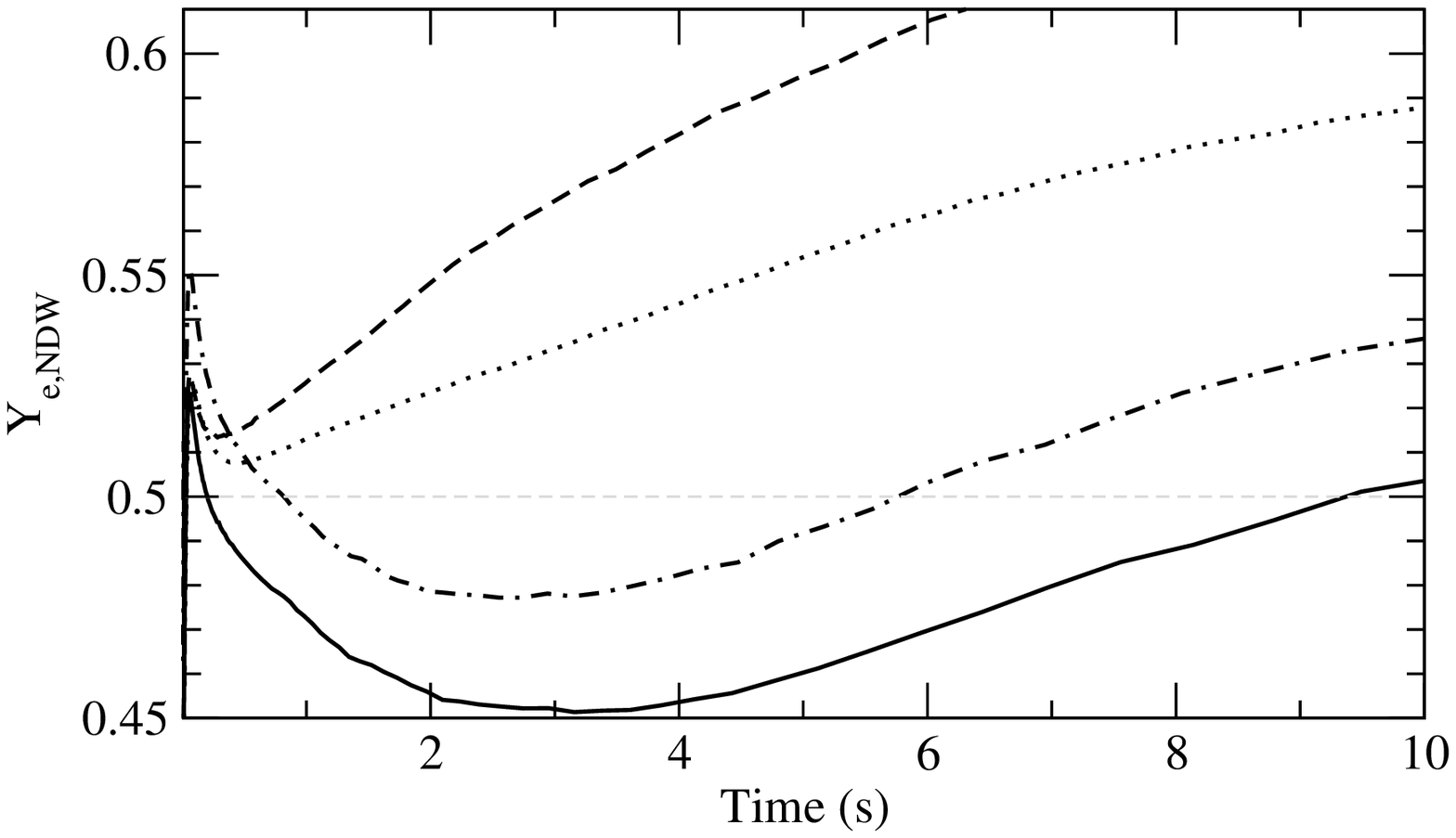}
\caption{
Top panel: Energy moments of the outgoing neutrino 
flux measured at the surface of the calculation
using the \cite{Bruenn85} approximation for the electron neutrino
and anti-neutrino capture rates on nucleons (black lines).  The
gray lines are for the fiducial model using the full capture 
cross-sections from \cite{Reddy98}.  There is little variation
for the $\bar \nu_e$ and $\nu_x$ energies between the two 
cross-section prescriptions, but there is a significant change
in the $\nu_e$ average energies.
Bottom panel: Predicted neutrino driven wind electron fraction
as a function of time.  The dotted line is from a PNS model using
the \cite{Bruenn85} rates, the solid line is for the \cite{Reddy98} 
rates including tensor polarization corrections and the mean fields
\citep{Horowitz03}, the dot-dashed line is a model using the \cite{Reddy98} 
rates without tensor polarization corrections and with mean fields, 
and the dashed line is a model using the \citet{Reddy98} rates with 
tensor polarization corrections but neglecting the effects of mean fields.  
Note that neutron richness is predicted from about 1.5 to 10 seconds when 
realistic kinematics is used in the capture rates, while the wind is 
predicted to be proton rich throughout when the effects of the neutron
and proton potentials are ignored.}
\label{fig:Average Energy Compare}
\end{center}
\end{figure}
  
To illustrate how more realistic rates affect the predicted neutrino 
properties, a model identical to the one described in section 
\ref{sec:PNSEvolution} was run, except that the nucleon capture 
rates were replaced with the \cite{Bruenn85} capture rates neglecting 
the nucleon potentials.  The luminosities as a function of time are 
shown in figure \ref{fig:Luminosity_compare}.  The changes in the 
luminosity are relatively small.  The most obvious difference is that 
the luminosities of all neutrino species asymptote to one another 
at late times, which is similar to the behavior seen in 
\cite{Fischer11}.  From one to ten seconds, there is a significantly
smaller difference between the luminosities than in the 
fiducial model.  Cooling via electron neutrinos and 
anti-neutrinos is also increased at late times, but the 
$\nu_x$ luminosity is virtually unchanged, as expected.  Before
1 s, the electron neutrino luminosity is reduced.  It is 
unlikely that this is significant, as the first approximately hundred 
milliseconds of these simulations are suspect for the reasons 
described above.  The reason for the early time decrease is less
clear.  It is unlikely that this is due to the effects of nuclear
interactions because the region where electron neutrinos decouple is
in the mantle which is at low density.

The evolution of the average neutrino energies are shown 
in figure \ref{fig:Average Energy Compare}.  There is little 
change between the models in the electron anti-neutrino, $\mu$, 
and $\tau$ neutrino average energies, but there is a significant
change in the electron neutrino average energies.  At early
times the average energy is reduced compared to the fiducial 
model and at late times it is increased.  The late-time 
convergence is easily explained by the argument given in the 
paragraphs above and by the arguments given in \cite{Fischer11}.  

This difference is important to the composition of the 
neutrino driven wind.  The electron fraction of the neutrino
driven wind can be estimated as \citep{Qian96} 
\be
\label{eq:Ye_NDW}
Y_{e,{\rm NDW}} \approx \left [1 + \frac{\dot N_{\bar \nu_e}
\langle \sigma(\epsilon)_{p,\bar\nu_e}\rangle}{\dot N_{\nu_e}\langle 
\sigma(\epsilon)_{n,\nu_e}\rangle}  \right]^{-1}
\ee  
where $\langle \sigma\rangle$ are the energy averaged cross-sections
for neutrino capture on nucleons, which are approximately proportional
to $\epsilon^2$.  Smaller relative $\nu_e$ average 
energies and lower de-leptonization rates lead to a lower electron
fraction in the wind.  

The evolution of the electron fraction in the neutrino driven wind 
calculated using equation \ref{eq:Ye_NDW} for both models, as well 
as a model that does not include weak magnetism corrections and a 
model that does not include mean field effects, but which do 
include full kinematics in the structure functions, is shown in the 
second panel of figure \ref{fig:Average Energy Compare}.
The capture rates for low densities given in \cite{Burrows06},
which include first order weak magnetism and recoil corrections,
have been used.  This was done to put the comparison between the
 models on even footing, although it is not necessarily 
consistent with the rates used inside the PNS itself.  The alpha
effect \citep{Fuller95} has also not been taken into account, which 
will push $Y_e$ closer to a half in both proton and neutron rich
conditions.  Energy moments of the neutrino flux are taken using 
the values at the surface of the computational domain, not at 
infinity.  The $\dot N_{\nu_e}/\dot N_{\bar\nu_e}$ term is increasing 
with time in both models (see figure \ref{fig:Luminosity_compare}),
which increases the electron fraction in the wind.  Note that once
neutrinos are free streaming, this term is invariant with radius.

With these assumptions, the fiducial model of section \ref{sec:PNSEvolution}
actually does result in a period of neutron richness in the wind, in 
contrast to the results of \cite{Huedepohl10} and \cite{Fischer11}.  
The wind is not very neutron rich ($Y_e \gtrsim 0.45$ at all times) and 
this change, by itself, would not result in substantial $r$-process 
nucleosynthesis in the standard neutrino driven wind where entropies 
are $\lesssim 150$ \citep{Roberts10}.  If for some reason
the entropy were higher though, the possibility of an $r$-process
remains. In contrast, the model that uses the rates of \cite{Bruenn85}
and the model using the \citet{Reddy98} rates without isospin dependent 
nuclear potentials consistent with the underlying equation of state 
results in a wind that is always proton-rich.  

To emphasize that this
result is mainly due to the reaction kinematics and not the inclusion 
of weak magnetism, models with the tensor polarization set to zero are 
also shown in the bottom panel of figure \ref{fig:Average Energy Compare}.
As is expected from the first order weak magnetism corrections given
in \cite{Horowitz02}, allowing for a tensor portion of the response
increase the difference between the electron neutrino and
anti-neutrino average energies.  This results in a lower electron
fraction in the case including the tensor polarization relative to
the case without.  Still, the change between these two models is only
a fraction of the change in the electron fraction when the \citet{Bruenn85}
rates are used.

Given the sensitivity to the neutrino interaction rates, it is
possible that further improvement of the treatment of electron neutrino and
anti-neutrino capture will alter this conclusion in one direction or
the other.  Because the asymmetry between the electron neutrino capture
rates depends on the value of the symmetry energy and the symmetry energy
varies with density \citep{Fattoyev10}, it may be that different nuclear 
equations of state alter the predicted neutron excess in the NDW.  
Variations of the calculation of the rates, such as incorporating 
the effect of correlations in the nuclear medium, can
significantly change the timescale of the neutrino emission \citep{Reddy99} 
and possibly the spectral properties.  Even if updated rates only change
the rates at high density, this may affect $\dot N_{\nu_e}/\dot N_{\bar\nu_e}$
and thereby change the properties of the wind.  Such extensions depend on the 
underlying equation of state, which is uncertain, and also on approximations
inherent to many-body theories of strongly interacting systems.  While 
this deserves further consideration, the results presented here seem to
indicate that effects due to kinematics, degeneracy and mean fields are
crucial.

It bears mentioning that the study of \citet{Huedepohl10} did allow 
for energy and momentum transfer between the nucleons and leptons 
\citep{Buras06}, but did not account for the difference between the 
neutron and proton mean field potentials.  Their rates were calculated within 
the random phase approximation of \citet{Burrows99}, which is an improvement over 
the mean field rates used in this work.  Additionally, weak magnetism corrections 
were approximately included in this study via the prescription of 
\citet{Buras06}.  Their average energies were further apart than when the 
\cite{Bruenn85} rates were used, but the difference was still not great 
enough to result in a wind with a neutron excess.  The average energies 
of the electron neutrinos and anti-neutrinos also asymptote to one another 
fairly rapidly in \citet{Huedepohl10}, in contrast to the present study.  
This is all reasonably consistent with the affect of neglecting the mean 
field potentials in the nucleon kinematics, but it is far from certain that
this is the main source of discrepancy.  Further exploration of why this 
work differs from the work of \citet{Huedepohl10} is surely warranted.

Even given these caveats, it is tantalizing that the wind 
is neutron rich in the fiducial model once again.  Extensions of the 
standard neutrino driven wind model which include heating from a source 
besides neutrinos can produce the $r$-process even for such modest neutron 
excesses \citep[][]{Suzuki06}. 

\subsection{Time Integrated Spectra}
\label{sec:Integrated Spectra}
\begin{figure}
\begin{center}
\leavevmode
\includegraphics[width=\columnwidth]{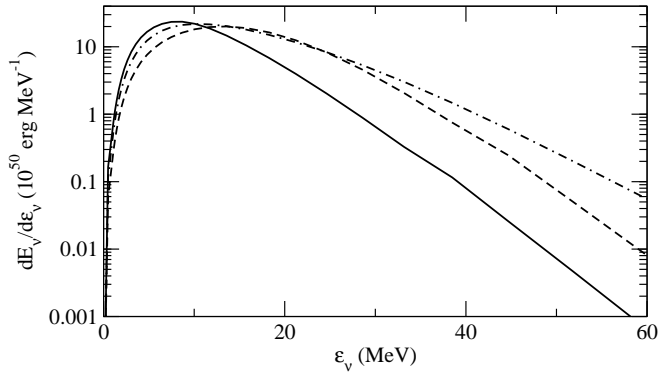}
\caption{Spectrum of integrated neutrino emission 
over the duration of the simulation.  Although
the peak of the $\bar \nu_e$ spectrum corresponds 
to the peak of the $\nu_x$ spectrum, the $\nu_x$
spectrum has a significantly harder tail.}
\label{fig:Integrated Spec}
\end{center}
\end{figure} 

In figure \ref{fig:Integrated Spec}, the 
integrated neutrino luminosity as a function of neutrino 
energy at infinity is shown for the model described in 
section \ref{sec:PNSEvolution}.  The time integrated 
average energies of the neutrinos are $\langle \epsilon_{\nu_e} 
\rangle = 8.3 \, {\rm MeV}$, $\langle \epsilon_{\bar \nu_e} 
\rangle = 12.2 \, {\rm MeV}$, and $\langle \epsilon_{\nu_x} 
\rangle = 11.1 \, {\rm MeV}$.  Although the $\mu$ and $\tau$ 
neutrinos are as hot or hotter than the electron anti-neutrinos 
at early times, the time integrated average 
is weighted more strongly towards late times so that they in fact 
have a somewhat lower average energy.

Time integrated neutrino spectra are interesting for both
nucleosynthesis via the $\nu$-process \citep{Heger05} and for
predictions of the diffuse supernova neutrino background
\citep{Ando04}.  The neutrinos are non-thermal and are not easily
described by an effective Fermi-Dirac distribution.  Given the
sensitivity of the $\nu$-process to the energy of the emitted
neutrinos (especially the energies above threshold), it seems that
calculations of neutrino-induced nucleosynthesis needs to be done with
more accurate neutrino spectra to check previous results in the
literature.  These integrated spectra are only approximate however
because a substantial fraction (20\%) of the neutrinos are emitted
during the first second of mantle contraction, this phase of evolution
contributes the majority of the high-energy tail, and the mantle
contraction phase is most sensitive to the approximate initial
conditions used.

\section{Conclusions}
\label{sec:conclusion}

A new code for following the evolution of PNSs has been described and
some first results obtained.  In section \ref{Sec:MomentTransport}, a
formalism for moment based neutrino transfer with variable Eddington
factors has been described, based on the work of \citet{Thorne81} and
\citet{Lindquist66}.  The framework is fully general relativistic and
is formulated in the rest frame of the fluid, which simplifies
calculation of the collision terms.  The code employs energy integrated
groups, rather than discrete energies, for solving the radiative
transfer problem.  This makes it well suited for dealing with problems
were thermodynamic equilibrium holds in large portions of the problem
domain and the distribution functions may contain sharp Fermi
surfaces.  A method for finding Eddington factors from a formal
solution to the static Lindquist equation was also
described. Additionally, general descriptions of the source terms for
absorption, scattering, and pair annihilation have been provided which
are consistent with the formalism, explicitly obey detailed balance,
and therefore naturally deal with the transition to equilibrium.  The
details of a fully implicit numerical implementation of these
transport equations alongside the equations of general relativistic
hydrodynamics were then described in section \ref{sec:Numerics}.

The results of a fiducial model of PNS cooling were presented in
section \ref{sec:PNSEvolution}.  The evolution proceeds similarly
to previous results in the literature in which a similar nuclear 
equation of state and neutrino opacities were used.  I have focused 
on the spectral properties of the emitted neutrinos, which were not 
well described by the formalism of \cite{Pons99}.  Similar behavior 
is found to other recent results in the literature: spectral softening
as a function of time and convergence of the $\bar \nu_e$ and 
$\nu_{\tau,\mu}$ luminosities after about two seconds of evolution
\citep{Fischer10,Huedepohl10}.  

In contrast to other recent studies \citep[c.f.][]{Fischer11} however,
the new studies show that the average energy of the electron neutrinos
does not converge to the average energies of the other neutrino
flavors at late times.  Additionally, the electron neutrinos are
significantly cooler than the anti-electron neutrinos for most of the
simulation.  This difference is likely due to the treatment of charged
current neutrino interactions, where a realistic
treatment of the nucleon kinematics including the nuclear potential 
is important (see section \ref{sec:Spectra Compare}).  The implications 
of this result for the electron fraction in the neutrino driven wind 
and possible $r$-process nucleosynthesis in this environment were 
discussed and warrant further exploration.

A quantitative comparison was also made between the results of an 
EFLD calculation of PNS evolution and evolution with the code 
described in this paper in section \ref{sec:EFLDComp}.  It was 
found that EFLD provides a good approximation to the total neutrino 
luminosity during periods in which the neutrino luminosity is dominated 
by emission from optically thick regions.  This approximation does 
break down in the optically thin regime as expected.  Additionally, 
it does a poor job of predicting the luminosities of individual 
neutrino flavors.

The most significant improvement which could be made to this 
work would be to improve the initial models.  This could be done
either by using a  more realistic post-core collapse initial 
models or updating the code to allow it to follow collapse and 
bounce itself, similar to \cite{Huedepohl10} and \cite{Fischer10}.
Such improvements will affect the early time evolution, but are
probably less important to the evolution of the PNS after one second.
Additionally, a more realistic equation of state that includes nuclei
at low densities \citep{Shen11} should be employed with consistent 
opacities.  

Of course, this work has also been limited to one dimension, which may 
be a gross, although necessary, oversimplification.  Convection, 
magnetic fields, and rotation may be central players in the evolution
of PNSs.  Approximate mixing length convection will be included in a 
subsequent version of the code.  In the future, this code will be 
applied to understanding the diffuse supernova neutrino background, 
predicting the affects of different prescriptions for the nuclear 
equation of state on PNS cooling, and investigating black hole 
formation.  


\begin{acknowledgements}
I gratefully acknowledge Sanjay Reddy, Vincenzo Cirigliano, and Gang Shen 
for useful discussions about this work and for help with the 
opacities and equation of state used in sections \ref{sec:PNSEvolution} and
\ref{sec:Spectra Compare}.  I also thank Stan Woosley for numerous 
useful discussions concerning this work and for a careful reading of the 
manuscript.  I acknowledge support from the University of California Office 
of the President (09-IR-07-117968-WOOS) and assistance during the early 
stages from an NNSA/DOE Stewardship Science Graduate Fellowship 
(DE-FC52-08NA28752).  This research has also been supported at
UCSC by the National Science Foundation (AST-0909129).
\end{acknowledgements}

\appendix

\section{Code Tests}
\label{sec:CodeTests}

\begin{figure}
\begin{center}
\leavevmode
\includegraphics[width=0.5\columnwidth]{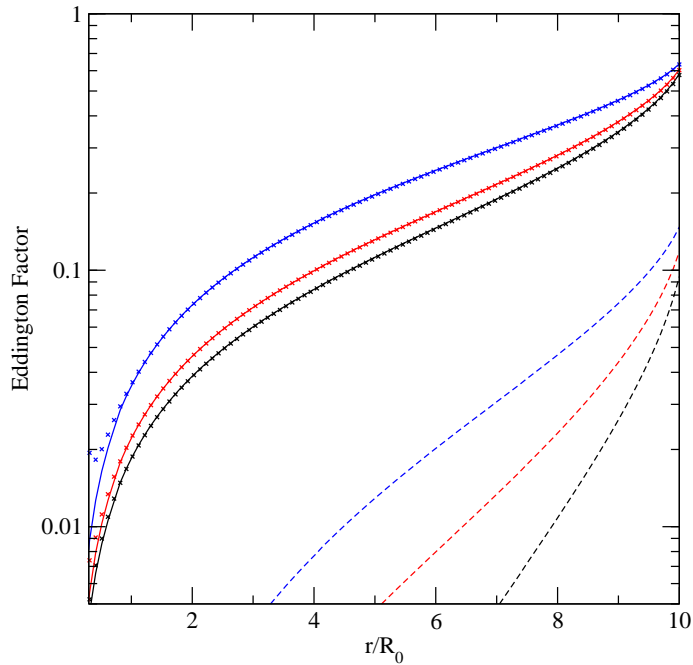}
\caption{Comparison of $H_g/E_g$ (crosses) and $g_1$ (solid lines) for 
a purely absorbing (black lines) and an isotropic scattering sphere 
(red lines) with total optical depth one and a first order scattering 
sphere (blue lines) with $\tau = 0.1$.  Aside from at the center, 
there is excellent agreement between the formal solution and the results 
of the moment calculation.  The dashed lines are the second Eddington 
factors, $g_2$, for the same models.}
\label{fig:Eddington Factor Comparison}
\end{center}
\end{figure} 

\begin{figure*}
\begin{center}
\leavevmode
\includegraphics[width=0.45\columnwidth]{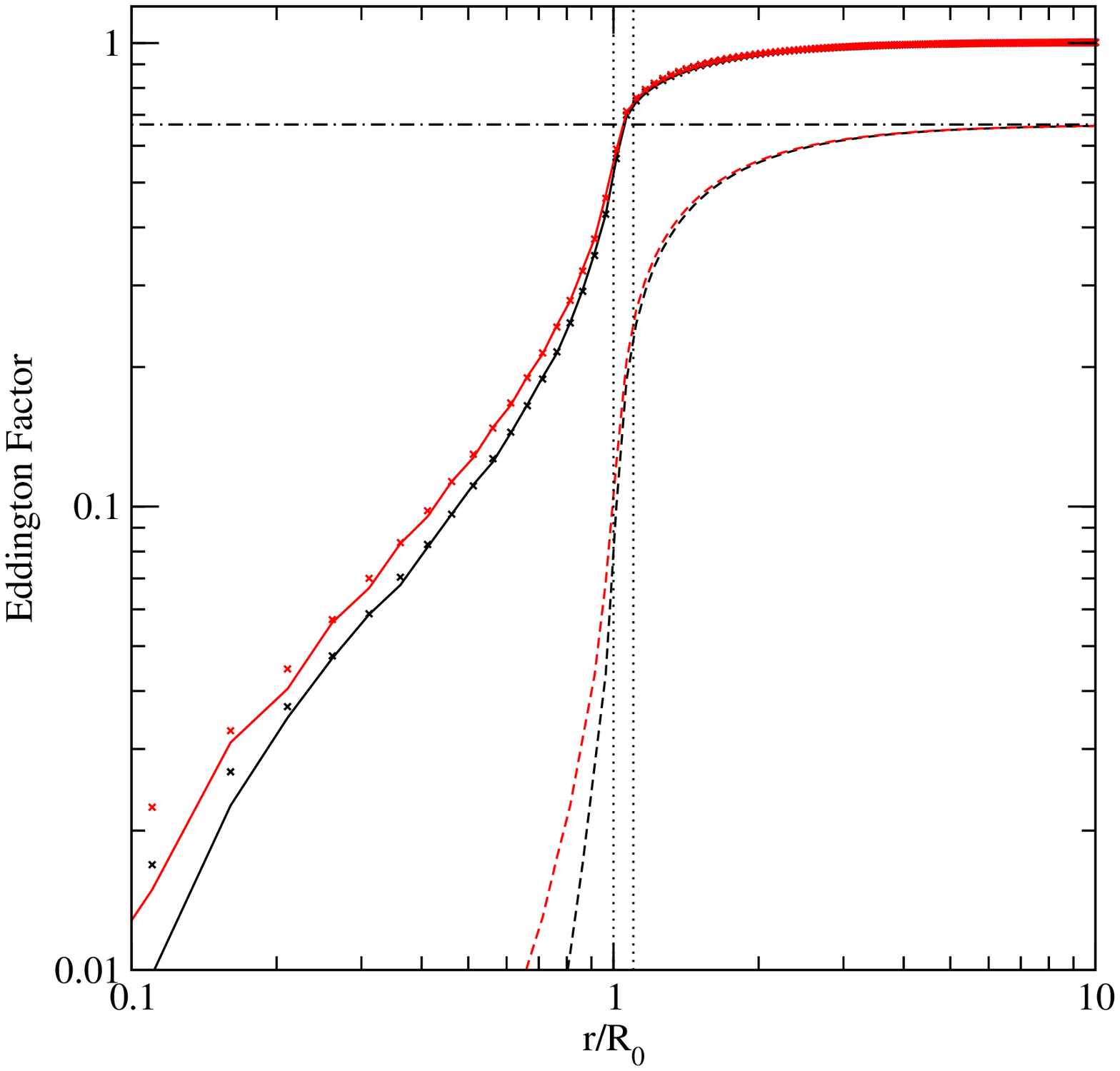}
\includegraphics[width=0.45\columnwidth]{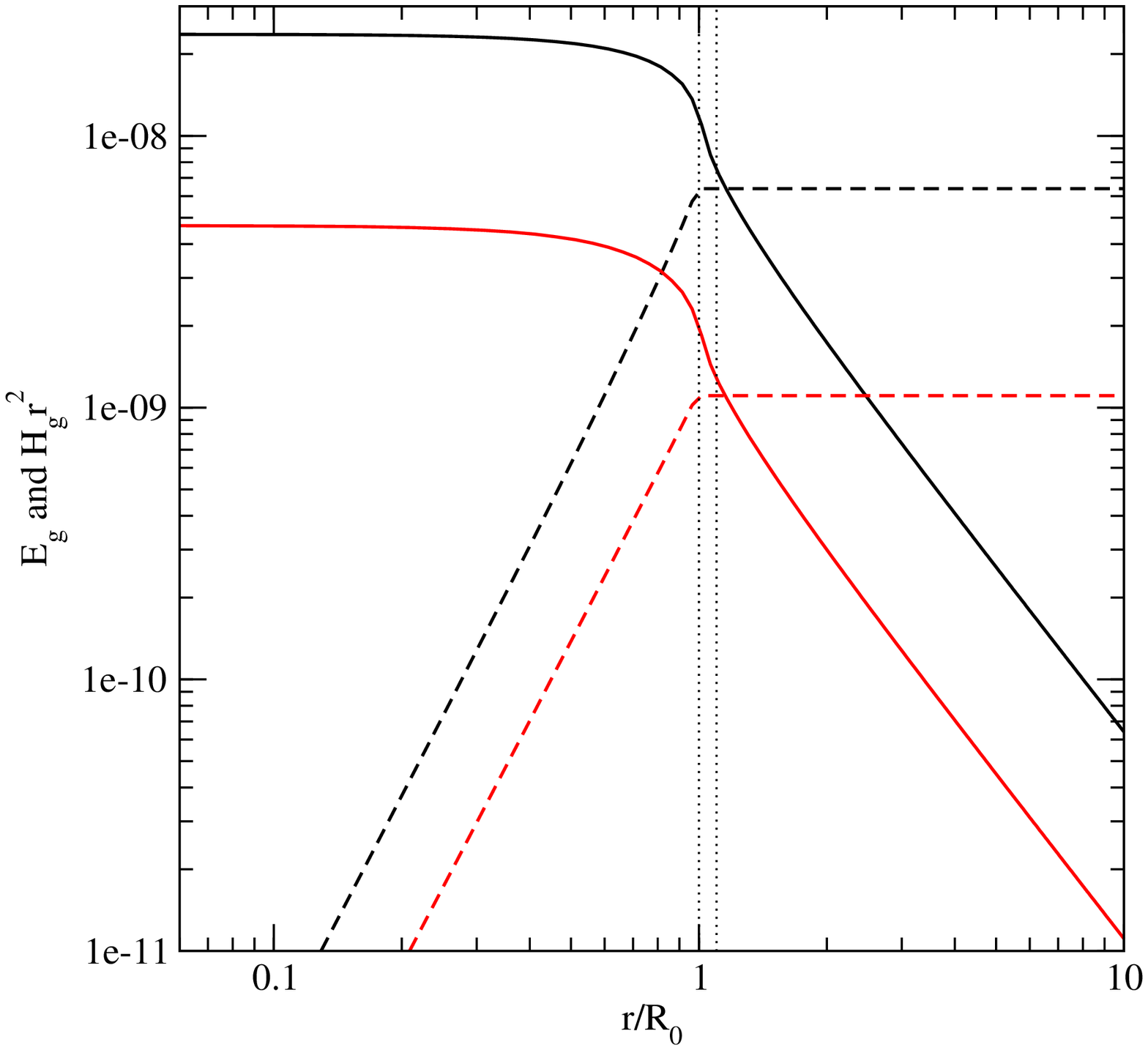}
\caption{ Top Panel: Eddington factors for radiation streaming from a 
homogenous sphere into free space.  The solid lines are the Eddington
factors $g_1$ for the purely absorptive sphere (black) and the isotropic 
scattering sphere (red).  The crosses show $H_g/E_g$.  The dashed lines 
are the second Eddington factor, $g_2$, for the same models.  The vertical
dotted lines show the radii at which the opaque sphere ends.  The horizontal 
dashed line shows the expected asymptotic value of $g_2$ for free streaming
radiation.
Bottom Panel: Properties of the radiation field as a function of radius.  The 
solid lines show the radiation energy density and the dashed lines show the 
luminosity per steradian, $r^2 H_g$. The colors are the same as in the top panel.
Once again, the vertical dotted line denotes the end of the opaque sphere.}
\label{fig:Sphere Decoupling edd facs}
\end{center}
\end{figure*}
 
Here I consider static transport through an homogeneous sphere with unit
radius, and unit optical depth.  A flat space-time is assumed.  
The first test performed is for consistency between the formal solution of the Boltzmann 
equation and the moment equations.  In addition to the factors $g_2$ and 
$g_3$, the formal Boltzmann solver can also calculate $g_1 \equiv w^1/w^0$ 
which should be equal to $H_g/E_g$.  For a purely absorptive atmosphere,
the formal solution is exact.  A comparison of $g_1$ and $H_g/E_g$ is 
shown in figure \ref{fig:Eddington Factor Comparison}.  For this calculation,
one hundred equally spaced radially zones and a grid of 150 tangent rays 
with impact parameters spaced equally in radius were used.  The calculation
was then evolved for ten units of time.  The differences between the two Eddington 
factors are negligible, aside from in the inner most zones.  This agreement 
does not depend strongly on the number of tangent rays employed.
The disagreement in the inner most regions is due to the small number of 
tangent rays which have impact parameters that are less than the radius 
at which the Eddington factor is calculated.  The distribution function 
is not well resolved and is therefore in error.  Such problems should not 
arise in the actual evolution of PNSs, as the inner most regions are 
generally opaque.

As a second test, a sphere which includes a scattering 
contribution to the opacity is considered.  In this case, the formal Boltzmann solver no longer 
gives an exact solution of the transport equation because of the 
approximate treatment of the scattering terms (see section 
\ref{sec:FormalSolution}).  The problem set up involves a sphere of optical 
depth one with $10\%$ of the opacity coming from absorption and $90\%$ 
coming from isotropic scattering. Using a similar numerical set up to the 
absorbing atmosphere gives the results shown in figure \ref{fig:Eddington 
Factor Comparison}.  Once again the deviation between the formal 
solution and the results of the moment calculation are small.  Above the 
two innermost zones, the maximum deviation of the two results for $g_1$ 
is less than $1\%$.  At radii greater than 8, the deviation is less than
$0.1\%$.  A second scattering test problem was run with the isotropic
scattering opacity, $\chi_0^s$, set to zero and the first order scattering
opacity, $\chi_1^s$, set to $90\%$ of the total opacity.  The sphere was 
assumed to have an optical depth $\tau = 0.1$.  The agreement between the 
moment calculation and the formal solution was found to be similar to the 
isotropic scattering case.

Tests similar to the homogeneous sphere tests of \citet{Rampp02} have also been performed.  
In these, a unit optical depth sphere of 
radius one is included inside a transparent region of radius ten, 
with a sharp transition region from the semi-opaque sphere to the
surrounding vacuum.  Such a scenario is similar to the neutrino 
decoupling region of PNSs, and is therefore an important test problem
for any neutrino transport code for PNS evolution.  The calculation domain is split up into
200 zones, equally spaced in radius and 301 a grid of 301 tangent 
is employed.  Only the radiation density, $E_g$ and flux, $H_g$, are
evolved.  The calculation is then run for fifty time steps, which
allows the calculation to relax to steady state and forget  the 
details of the initial conditions of the radiation field.  Once
again, one calculation was run with a purely absorbing opacity and 
a second was run with $10\%$ absorbing opacity and $90\%$ isotropic 
scattering opacity.  

The final Eddington factors and radiation field 
for these calculations are shown in figure \ref{fig:Sphere Decoupling edd facs}.
For the purely absorbing calculation, the formal solution of the 
Boltzmann equation is exact.  In the inner most zones, the distribution 
function is under resolved in angle due to the small number of tangent 
rays which pass through this region.  This is not a problem for PNS 
simulations, as the optical depth in the interior is always much larger 
than one.  Therefore, the Eddington factors are close to zero and have 
negligible affects on the moment transport solution.   The deviation of 
the moment solution, outside the inner most region, from the formal 
solution is less than $1\%$.  In the decoupling region the agreement 
is excellent.  Additionally both $g_1$ and $g_2$ asymptote to their 
expected values for free streaming radiation far from the core. 

\section{Equilibrium Flux Limited Diffusion}
\label{sec:EFLD}
In the diffusion approximation, the distribution function is 
approximated by $f(\omega,\mu) = f_0(\omega) + f_1(\omega)\mu$.  
This results in $g_2 = g_3 = 0$.  The time dependence is then
dropped and compression terms are then dropped from the first 
moment equation.  Ignoring the terms containing derivatives 
with respect to energy (which will drop out in the end of our
analysis anyway), I find
\be
\Gamma \left(\pd{w^0}{r} + 4 \pd{\phi}{r} w^0 \right) = 3 s^1
\ee 
If the redshifted frequency is denoted as $\nu=\omega e^\phi$, 
the redshifted source function is 
\be
s^1 = - w_1 D^{-1}(\nu e^{-\phi})
\ee
which gives the redshifted neutrino flux per energy
\be
w_1 = -\Gamma \frac{D(\nu e^{-\phi})}{3} \left(\pd{w^0}{r} 
+ 4 \pd{\phi}{r} w^0 \right)
\ee
where $D(\omega) = (1/\lambda_a^* + \chi_0^s - \chi_1^s/3)^{-1}$.
This amounts to assuming that the neutrino flux instantaneously 
equilibrates to the gradients in the neutrino number density.  Because
a number of terms have been dropped, this expression does not guarantee
that $w_1\leq w_0$, which means that it is possible for the neutrino 
fluxes to violate causality \citep{Levermore81}.  This problem is usually
circumvented by introducing a flux limiter, which is a correction to 
the diffusion coefficient which depends on $\xi = w_1/w_0$ and serves
to keep the fluxes causal.  In section \ref{sec:EFLDComp}, the flux
limiter of \cite{Levermore81} has been used, but it was found in 
\cite{Pons99} that the results of EFLD calculations are reasonably
insensitive to the choice of flux limiter.  

The total energy flux is given by
\be
H = \int d\omega w^1 = - e^{\phi}\Gamma \int d\omega  \frac{D(\nu e^{-\phi})}
{3} \left[\pd{w^0}{r} - e^{4\phi} \pd{e^{-4\phi}}{r} w^0 
\right]
\ee
The equilibrium portion of the EFLD approximation constitutes 
assuming that $f_0(\omega) = f_{{\rm eq}}(\omega,T,\mu)$.  Using 
the equilibrium expression for $w_0$ yields
\be
H = -\frac{\Gamma}{ 6 \pi^2 }\int d\omega D(\omega) \omega^3 
\pd{f_{{\rm eq}}(\omega)}{r}
\ee
The radial derivative is then given by
\be
\pd{f_{{\rm eq}}(\nu e^{-\phi})}{r} =
\left[\frac{\omega}{T} \pd{T e^\phi}{r} + Te^{\phi} \pd{\eta}{r} \right]
\frac{f_{{\rm eq}}(\omega)(1-f_{{\rm eq}}(\omega))}{Te^\phi}
\ee
So that the energy flux is given by
\be
H = -\frac{\Gamma e^{-\phi} T^3}{ 6 \pi^2 }\left[D_4 \pd{T e^\phi}{r}
+ D_3 T e^{\phi} \pd{\eta}{r} \right],  
\ee
where the energy integrated diffusion coefficients are defined as 
\be
D_n = \int_0^\infty dx x^n D(xT) f_{{\rm eq}}(xT)(1-f_{{\rm eq}}(xT)).
\ee
The number flux equation can easily be determined from the energy 
flux equation.  This gives
\be
F = -\frac{\Gamma e^{ -\phi} T^2}{ 6 \pi^2 }\left[D_3 \pd{T e^\phi}{r}
+ D_2 T e^{\phi} \pd{\eta}{r} \right].
\ee
These expressions agree with the results of \cite{Pons99}.

Now all that is left is to describe the evolution of the underlying 
medium.  Using equation \ref{eq:InternalEnergy}, the evolution of the 
total internal energy of the medium including neutrinos is found to be
\be
\pd{\epsilon}{t} 
+ e^{\phi}\Theta \frac{p}{n}
+ e^{-\phi}\pd{}{a} \left( \sum_{s \, = \atop \{\nu_e,\cdots\}}
4 \pi r^2 e^{2 \phi} H_{s} \right) = 0
\ee
Using equation \ref{eq:LeptonFraction}, the evolution of total 
lepton number is found to be
\be
\pd{Y_L}{t} + \pd{}{a}\left(4\pi r^2 e^{\phi} 
\left[F_{\nu_e}-F_{\bar \nu_e}\right]\right) = 0.
\ee

\end{document}